\documentclass[iop,apj]{emulateapj}
\usepackage{morefloats}

\shorttitle{V4641 Sgr: Activity in Quiescence \& Improved Masses}

\begin{document}

\title{The Black Hole Binary V4641 Sagitarii:\\Activity in Quiescence \& Improved Mass Determinations}

\author{Rachel K.D. MacDonald, Charles D. Bailyn, Michelle Buxton, Andrew G. Cantrell\altaffilmark{1}, Ritaban Chatterjee\altaffilmark{2}, Ross Kennedy-Shaffer\altaffilmark{3}}
\affil{Department of Astronomy, Yale University, PO Box 208101, New Haven, CT 06520-8101, USA}
\email{rachel.macdonald@yale.edu}
\altaffiltext{1}{The Blake School, 511 Kenwood Pkwy, Minneapolis, MN 55403, USA}
\altaffiltext{2}{Department of Physics, Presidency University, 86/1 College Street, Kolkata-700073, WB, India}
\altaffiltext{3}{Hunter College High School, 71 E 94th St, New York, NY 10128, USA}

\author{Jerome A. Orosz}
\affil{Department of Astronomy, San Diego State University, San Diego, CA 92182, USA}
\and

\author{Craig B. Markwardt, Jean H. Swank}
\affil{Astrophysics Science Division, Goddard Space Flight Center, NASA, Greenbelt, MD 20771, USA}

\begin{abstract}
We examine $\sim$10 years of photometric data and find that the black hole X-ray binary V4641 Sgr has two optical states, passive and active, during X-ray quiescence. 
The passive state is dominated by ellipsoidal variations and is stable in the shape and variability of the light curve. The active state is brighter and more variable. 
Emission during the active state varies over the course of the orbital period and is redder than the companion star. 
These optical/infrared states last for weeks or months. V4641 Sgr spends approximately 85\% of X-ray quiescence in the passive state and 15\% in the active. 
We analyze passive colors and spectroscopy of V4641 Sgr and show that they are consistent with a reddened B9III star (with $E(\bv) = 0.37 \pm 0.19$) with little or no contribution from the accretion disk.
We use X-ray observations with an updated ephemeris to place an upper limit on the duration of an X-ray eclipse of $<8.3\degr$ in phase ($\sim$1.6 hours).  
High resolution spectroscopy yields a greatly improved measurement of the rotational velocity of the companion star of $V_{\rm rot}\sin i=100.9\pm 0.8$ km s$^{-1}$. 
We fit ellipsoidal models to the passive state data and find an inclination angle of $i = 72.3 \pm 4.1 \degr$, a mass ratio of $Q = 2.2\pm 0.2$, and component masses for the system of $M_{BH} = 6.4\pm 0.6$ M$_{\sun}$ and $M_{2} = 2.9 \pm 0.4$ M$_{\sun}$. Using these values we calculate an updated distance to V4641 Sgr of $6.2 \pm 0.7$ kpc.
\end{abstract}


\section{Introduction}
X-ray binaries (XRBs) are binary systems in which a companion star is transferring mass to a compact object, leading to accretion-driven X-ray outbursts. XRBs can be classified into two broad categories: high-mass and low-mass, referring to the ratio of the mass of the companion star to the compact object. In general, high-mass XRBs transfer mass to their compact object via a strong stellar wind, and low-mass XRBs transfer mass when the companion star overflows its Roche lobe. The compact object in the system can be either a neutron star or a black hole. One subset of XRBs is referred to as soft X-ray transients or simply X-ray transients; these can be either high-mass or low-mass. A typical soft X-ray transient has short X-ray outbursts driven by episodic accretion, which rise quickly and decay exponentially, and longer X-ray quiescent periods during which the luminosity is dominated by the companion star. X-ray outbursts in black hole XRBs are characterized by several different X-ray ``states'', which are differentiated by their spectral and timing properties.

V4641 Sgr is a low-mass X-ray binary (LMXB) and a soft X-ray transient, with a B9III companion star. 
It was discovered as an X-ray source independently in February of 1999 by both \textit{BeppoSAX} \citep{int_zand_sax_1999,int_zand_bepposax_2000} and \textit{Rossi X-ray Timing Explorer} \citep{markwardt_xte_1999}. It was discovered in radio in September of 1999 \citep{hjellming_light_2000}, shortly after undergoing two X-ray flares. Although confused for a time with the variable star GM Sgr, its true optical counterpart was determined to be a variable star discovered by \citet{goranskij_variable_1978}, now designated V4641 Sgr \citep{orosz_black_2001}. 

V4641 Sgr is unusual in several characteristics, including its companion star and its X-ray outbursts. Its companion star is the most massive, brightest, and bluest of the galactic LMXB companion stars. This has led to the system being classified as a high mass X-ray binary by some \citep[for example,][]{chaty_optical_2003, revnivtsev_super-eddington_2002, pandey_low-frequency_2007}. However, since the mass transfer occurs via Roche lobe overflow, and since the companion star is not more massive than the black hole in the system, we maintain the label of LMXB in this work. The X-ray outbursts observed in this system do not follow the standard soft X-ray transient fast-rise-exponential-decay pattern. Rather they are more like flares, with a fast rise and an equally fast decay --- for example, an X-ray outburst in 1999 September for V4641 Sgr rose to 12 crab in 8 hours and faded to 0.1 crab in less than 2 hours \citep{hjellming_light_2000}. No good explanation for this incongruous outburst behavior has yet been found.

Although known primarily for their different X-ray states, black hole XRBs also exhibit variability in the optical and infrared. Optical or infrared data taken during X-ray quiescence is often used to determine binary parameters of the system, such as mass function, mass ratio, and inclination angle, and then to determine the mass of the compact object in the system \citep[and references therein]{orosz_black_2001,kreidberg_mass_2012}. This analysis frequently assumes that if a system is in X-ray quiescence it is unchanging at all wavelengths --- that is, that the only variations in the system are due to orbital changes in viewing angle. 

Studies have shown that this is not always true \citep[see][and references therein]{Cantrell2008, Cantrell2010}, and that many systems display variability which cannot be explained by the ellipsoidal variations alone. In particular, \citet{Cantrell2008} showed that A0620-00 actually has three distinct optical states in which it can exist (designated ``active'', ``passive'', and ``loop''), all while being quiescent in X-rays. The optical passive state is when variations are solely due to orbital changes, while the active and loop states describe periods when there is a heightened accretion flow producing brighter and more variable emission.

The next logical issue is to determine whether A0620-00 is unique in having multiple optical states in quiescence, or whether this is commonplace among X-ray binaries. Here we begin to answer this question by examining one of the optically-brightest quiescent X-ray binary systems, V4641 Sgr. We find, by examining approximately ten years of photometric data, that this source also exhibits passive and active states during X-ray quiescence. The passive state is dominated by the bright companion star, with little contribution from the accretion disk, while the active state shows more emission and more variability, presumably due to an increase in the accretion flow in the system. In \S \ref{obssect} we describe our observations, in \S \ref{activepassivesect} we define the different optical states and discuss their properties in V4641 Sgr, in \S \ref{elcsect} we determine the inclination angle of the system and the component masses by modeling the system using only optically passive data, and in \S \ref{discussconcludesect} we discuss our results and offer some conclusions.

\section{Observations}\label{obssect}
\subsection{Photometry}
We have optical and near-IR photometry from the ANDICAM instrument on the 1.0m \& 1.3m telescopes operated by the SMARTS consortium, located at the Cerro Tololo Inter-American Observatory \citep{buxton_optical_2012}. The instrument was located on the 1.0m telescope prior to 2003, and was then moved to the 1.3m telescope, where it is still operating to the present day. Observations were taken every night or two from 2001 to 2010, usually from March until October, although this varied from season to season. Details of the observation dates can be seen in Table \ref{coveragedatestable}. Most of the time observations were in six bands ($B$, $V$, $I$, $J$, $H$, \& $K$), although occasionally due to instrumental problems only optical or only infrared data were taken. The source was only observed in $V$, $I$, $J$, \& $K$ during 2001, and was not observed at all in 2006. During late June 2010, the source was observed for several hours at a time on three consecutive nights. 

\begin{deluxetable}{ccc}
\tablewidth{0pt}
\tablecaption{Observing seasons for V4641 Sgr photometry \label{coveragedatestable}}
\tablehead{\colhead{Year} & \colhead{Optical Data} & \colhead{IR Data} }
\startdata
2001 & Mar 27 -- Oct 12 & Mar 27 -- Sep 5 \\
2002 & Feb 18 -- May 30 & Feb 22 -- Sep 25 \\
2003 & Apr 8 -- Nov 14 & Apr 8 -- Nov 14 \\
2004 & Mar 17 -- Oct 5 & Mar 23 -- Oct 31 \\
2005 & Oct 8 -- Nov 5 & Feb 12 -- Nov 4 \\
2007 & May 5 -- Sep 30 & May 5 -- Sep 29 \\
2008 & Feb 29 -- Nov 9 & Feb 29 -- Nov 9 \\
2009 & Feb 23 -- Sep 5 & Feb 22 -- Sep 1 \\
2010 & Feb 14 -- Oct 31 & Feb 14 -- Oct 31 
\enddata
\end{deluxetable}

Some images were rejected based on visual inspection. For the rest of the data, three field stars were used to check for bad nights: a night was rejected if the three stars' raw magnitudes were, on average, more than 1.5 standard deviations fainter than their mean values. A few observations were also removed from this dataset upon discovering that the source had gone into X-ray outburst (for example, 2003 August 5-8, as reported in \citet{2003ATel..170....1B}, \citet{2003ATel..171....1B}, and \citet{2003ATel..172....1R}). All photometric data presented here were obtained while the source was quiescent in X-rays. 

Individual observations of V4641 Sgr were taken in the following manner. For the optical bands, three exposures in each band were taken during each observation. Infrared data were taken in sets of five dithered exposures in each band. In June, 2010, when we observed several hours in a row on three nights, data were again taken in sets of three exposures for the optical and five exposures for the infrared.

All data reduction and photometry was done using standard IRAF\footnote{IRAF is distributed by the National Optical Astronomy Observatory, which is operated by the Association of Universities for Research in Astronomy (AURA) under cooperative agreement with the National Science Foundation.} tasks. 
For optical data, each exposure was reduced and had photometry done separately. Photometric data from each set of three exposures in a band were then averaged to produce one data point per band per night/observation. For infrared data, each set of five dithered exposures was sky-subtracted, shifted to a common coordinate system, and then summed to produce one exposure for each band each night. Photometry was done on this combined exposure to produce one data point per band per night/observation.
In June 2010 when multiple observations were taken throughout several nights, both optical and infrared exposures were obtained and processed in the same manner as described above.
Altogether this gives 565 points in $B$, 704 in $V$, 671 in $I$, 695 in $J$, 591 in $H$, and 648 in $K$. 

\begin{figure*}[t]
	\epsscale{1.05}
	\plottwo{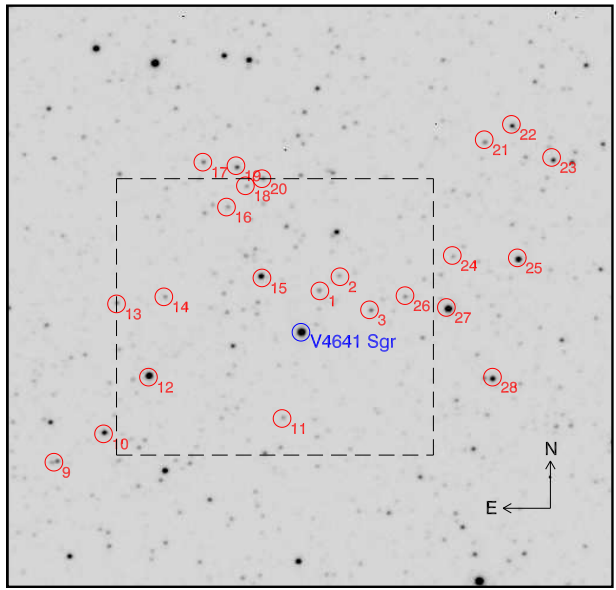}{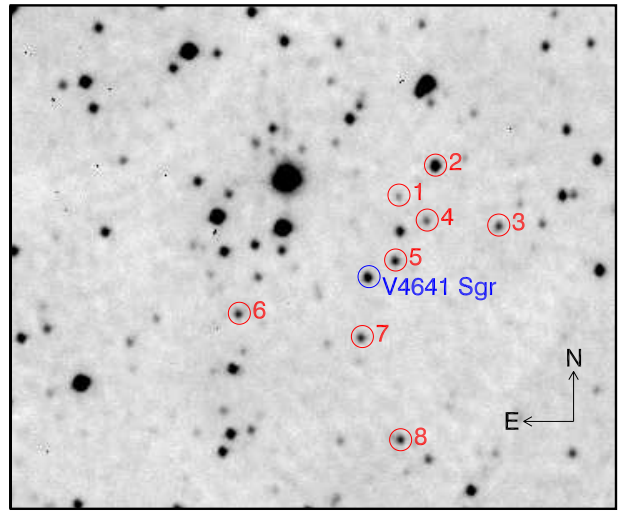}
	\caption{Finding charts showing field near V4641 Sgr in both optical and infrared bands. \textit{Left panel}: Optical ($B$ band, here) finding chart, showing V4641 Sgr and comparison stars used for optical differential photometry. Box outlined with dashed line shows approximate size and location of the IR field (shown in the other panel). \textit{Right panel}: Infrared ($K$ band, here) finding chart, showing V4641 Sgr and comparison stars used for IR differential photometry. \label{fig:findcharts}}
\end{figure*}

Optical differential photometry was calibrated using observations of standard stars \citep{landolt_ubvri_1992} taken on the same instrument by the SMARTS consortium. For 270 photometric observations taken between 2003 and 2005, calibrated magnitudes for V4641 Sgr in each filter were calculated using photometric zero points, color correction terms, and extinction correction terms derived from these standard star observations. Using these 270 observations, an average calibration factor was calculated for the differential photometry in each filter, and this factor was then applied to all observations of V4641 Sgr to convert them to magnitudes in the standard system. The infrared differential photometry was calibrated by deriving an average calibration factor from the known 2MASS magnitudes of six of the IR comparison stars. Finding charts showing V4641 Sgr and the comparison stars used for differential photometry and calibration are shown in Figure \ref{fig:findcharts}. Coordinates and magnitudes of comparison stars are shown in 
Table \ref{table:compstars}.

\begin{deluxetable*}{cccccc}
\tablewidth{0pt}
\tablecaption{Magnitudes of comparison stars \label{table:compstars}}
\tablehead{
\multicolumn{6}{c}{Optical Comparison Stars} \\
\hline \\[-4pt]
\colhead{Star} & \colhead{RA} & \colhead{Dec} & \colhead{$B$} & \colhead{$V$} & \colhead{$I$} \\
 & \colhead{(J2000)} & \colhead{(J2000)} & \colhead{(mag)} & \colhead{(mag)} & \colhead{(mag)}}
\startdata
1 & 18:19:21.0 & -25:24:10.0\tablenotemark{a} & 17.15 $\pm$ 0.16 & 16.24 $\pm$ 0.17 & 15.26 $\pm$ 0.16 \\
2 & 18:19:20.8 & -25:24:07.7\tablenotemark{b} & 17.72 $\pm$ 0.20 & 16.05 $\pm$ 0.17 & 14.36 $\pm$ 0.14 \\
3 & 18:19:20.0 & -25:24:18.3\tablenotemark{b} & 16.86 $\pm$ 0.14 & 15.82 $\pm$ 0.19 & 14.71 $\pm$ 0.15 \\
9 & 18:19:27.2 & -25:25:08.4\tablenotemark{a} & 16.47 $\pm$ 0.16 & 14.50 $\pm$ 0.17 & 11.53 $\pm$ 0.17 \\
10 & 18:19:26.2 & -25:24:59.3\tablenotemark{a} & 15.81 $\pm$ 0.13 & 14.42 $\pm$ 0.16 & 13.01 $\pm$ 0.13 \\
11 & 18:19:22.0 & -25:24:52.9\tablenotemark{a} & 17.54 $\pm$ 0.21 & 16.54 $\pm$ 0.23 & 15.49 $\pm$ 0.20 \\
12 & 18:19:25.1 & -25:24:37.4\tablenotemark{a} & 14.62 $\pm$ 0.12 & 13.41 $\pm$ 0.15 & 12.17 $\pm$ 0.13 \\
13 & 18:19:25.9 & -25:24:18.2\tablenotemark{a} & 17.14 $\pm$ 0.15 & 15.80 $\pm$ 0.16 & 14.40 $\pm$ 0.14 \\
14 & 18:19:24.9 & -25:24:15.2\tablenotemark{a} & 17.33 $\pm$ 0.17 & 15.96 $\pm$ 0.21 & 14.52 $\pm$ 0.15 \\
15 & 18:19:22.6 & -25:24:07.5\tablenotemark{a} & 15.53 $\pm$ 0.14 & 13.56 $\pm$ 0.16 & 11.11 $\pm$ 0.15 \\
16 & 18:19:23.5 & -25:23:47.2\tablenotemark{b} & 17.38 $\pm$ 0.21 & 16.15 $\pm$ 0.21 & 14.65 $\pm$ 0.38 \\
17 & 18:19:24.1 & -25:23:33.3\tablenotemark{b} & 17.03 $\pm$ 0.19 & 15.39 $\pm$ 0.19 & 13.68 $\pm$ 0.16 \\
18 & 18:19:22.9 & -25:23:40.5\tablenotemark{b} & 17.66 $\pm$ 0.19 & 16.04 $\pm$ 0.18 & 14.44 $\pm$ 0.15 \\
19 & 18:19:23.1 & -25:23:34.6\tablenotemark{b} & 16.50 $\pm$ 0.14 & 14.95 $\pm$ 0.17 & 13.34 $\pm$ 0.14 \\
20 & 18:19:22.6 & -25:23:38.0\tablenotemark{b} & 17.01 $\pm$ 0.16 & 15.56 $\pm$ 0.16 & 14.14 $\pm$ 0.14 \\
21 & 18:19:17.5 & -25:23:25.0\tablenotemark{a} & 17.11 $\pm$ 0.17 & 15.25 $\pm$ 0.18 & 13.29 $\pm$ 0.15 \\
22 & 18:19:16.9 & -25:23:19.3\tablenotemark{a} & 16.11 $\pm$ 0.14 & 14.84 $\pm$ 0.17 & 13.56 $\pm$ 0.14 \\
23 & 18:19:16.0 & -25:23:27.4\tablenotemark{a} & 16.29 $\pm$ 0.14 & 15.28 $\pm$ 0.17 & 14.26 $\pm$ 0.14 \\
24 & 18:19:18.2 & -25:24:00.0\tablenotemark{a} & 17.56 $\pm$ 0.17 & 16.59 $\pm$ 0.18 & 15.53 $\pm$ 0.17 \\
25 & 18:19:16.7 & -25:24:02.2\tablenotemark{a} & 15.29 $\pm$ 0.13 & 14.54 $\pm$ 0.16 & 13.70 $\pm$ 0.14 \\
26 & 18:19:19.2 & -25:24:13.6\tablenotemark{b} & 17.24 $\pm$ 0.16 & 15.86 $\pm$ 0.20 & 14.35 $\pm$ 0.15 \\
27 & 18:19:18.3 & -25:24:15.9\tablenotemark{a} & 14.52 $\pm$ 0.12 & 14.07 $\pm$ 0.15 & 13.64 $\pm$ 0.13 \\
28 & 18:19:17.2 & -25:24:38.1\tablenotemark{a} & 15.54 $\pm$ 0.14 & 13.86 $\pm$ 0.16 & 11.94 $\pm$ 0.20 \\
\hline \hline \\[-6pt]
\multicolumn{6}{c}{Infrared Comparison Stars} \\
 \hline \\[-4pt]
\colhead{Star} & \colhead{RA} & \colhead{Dec} & \colhead{$J$} & \colhead{$H$} & \colhead{$K$} \\
 & \colhead{(J2000)} & \colhead{(J2000)} & \colhead{(mag)} & \colhead{(mag)} & \colhead{(mag)} \\[2pt]
 \hline \\[-4pt]
1 & 18:19:21.0 & -25:24:10.0\tablenotemark{a} & 14.73 $\pm$ 0.33 & 14.34 $\pm$ 0.29 & 14.26 $\pm$ 0.22 \\
2 & 18:19:20.8 & -25:24:07.7\tablenotemark{b} & 13.08 $\pm$ 0.24 & 12.33 $\pm$ 0.19 & 12.16 $\pm$ 0.20 \\
3 & 18:19:20.0 & -25:24:18.3\tablenotemark{b} & 13.97 $\pm$ 0.27 & 13.50 $\pm$ 0.22 & 13.37 $\pm$ 0.17 \\
4 & 18:19:20.9 & -25:24:16.7\tablenotemark{b} & 14.56 $\pm$ 0.27 & 13.88 $\pm$ 0.21 & 13.75 $\pm$ 0.19 \\
5 & 18:19:21.3 & -25:24:23.3\tablenotemark{b} & 13.97 $\pm$ 0.28 & 13.28 $\pm$ 0.22 & 13.14 $\pm$ 0.28 \\
6 & 18:19:23.2 & -25:24:30.8\tablenotemark{b} & 13.85 $\pm$ 0.25 & 13.44 $\pm$ 0.21 & 13.36 $\pm$ 0.25 \\
7 & 18:19:21.7 & -25:24:35.7\tablenotemark{b} & 14.39 $\pm$ 0.27 & 13.61 $\pm$ 0.21 & 13.43 $\pm$ 0.24 \\
8 & 18:19:21.3 & -25:24:53.0\tablenotemark{b} & 14.07 $\pm$ 0.25 & 13.41 $\pm$ 0.21 & 13.26 $\pm$ 0.23 
\enddata
\tablecomments{
Star numbers are in accordance with the finding charts in Figure \ref{fig:findcharts}. 
Our infrared magnitudes are consistent with the 2MASS catalog within our photometric errors.}
\tablenotetext{a}{RA \& Dec are from the USNO-B1.0 Catalog.}
\tablenotetext{b}{RA \& Dec are from the 2MASS Point Source Catalog.}
\end{deluxetable*}

\begin{figure}[b]
	\epsscale{1.15}
	\plotone{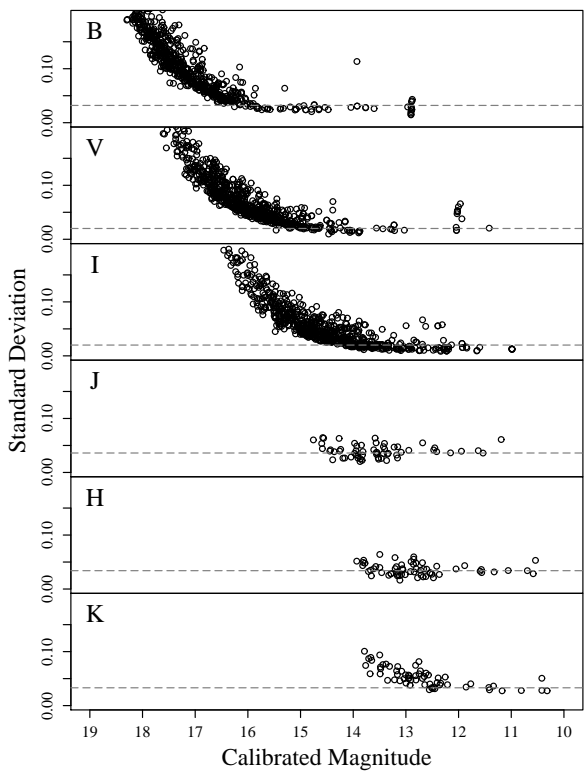}
	\caption{Standard deviation of differential magnitudes plotted against calibrated magnitudes, for field stars in each band. A fourth-order polynomial was fit to the distribution, and the minimum of that polynomial was extended out to brighter magnitudes in order to determine the typical error for the differential photometry. Horizontal lines indicate these minimum values for each band. \label{fig:photerrors}}
\end{figure}

Differential photometric errors were calculated from field stars by plotting the standard deviations of their differential photometry (calculated over $\sim300$ observations) against their calibrated magnitudes. We fit a fourth-order polynomial to these data, found the minimum of the curve, and extended that constant numerical value out to brighter magnitudes as the error on the differential photometry. Typical differential photometric errors for the brighter magnitudes were: 0.032 mag in $B$, 0.020 mag in $V$, 0.020 mag in $I$, 0.036 mag in $J$, 0.034 mag in $H$, and 0.033 mag in $K$. See Figure \ref{fig:photerrors} for the plots of standard deviation against calibrated magnitude and for the errors on fainter magnitudes. 

Calibration errors were calculated by taking the standard error of the mean of the comparison stars' calibrated magnitudes over the number of photometric nights used for the calibration. These calculated calibration errors averaged 0.006 mag for the optical bands and 0.025 mag for the infrared. There are undoubtedly systematic errors involved in the photometric calibration, concerning the exact match between our filters and those of the standard system. We took these systematic errors to be a few hundredths of a magnitude, giving final calibration errors of $\approx 0.02$ mag for the optical bands and $\approx 0.04$ mag for the infrared. Total photometric errors on the calibrated magnitudes in each band are therefore: 0.04 mag for $B$, 0.03 mag for $V$ and $I$, and 0.05 mag for $J$, $H$, and $K$.

\subsection{Spectroscopy}
\begin{figure}[t]
	\epsscale{1.15}
	\plotone{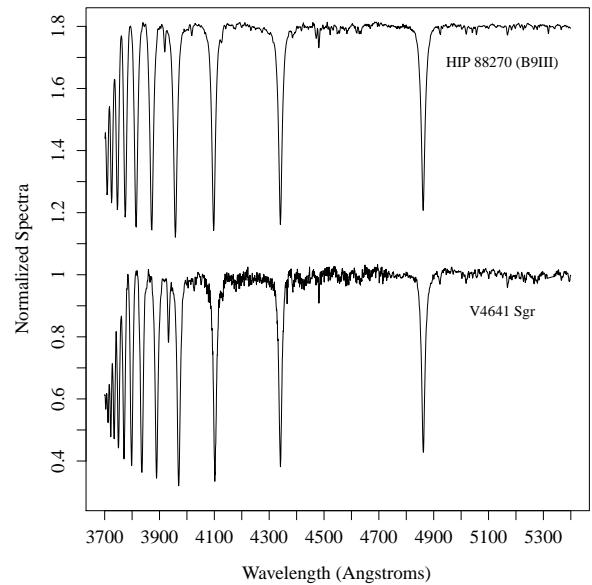}
	\caption{Normalized spectra taken on the SMARTS 1.5m telescope. Shown are V4641 Sgr and a comparison B9III star, HIP 88270 (offset by +0.8 units).  \label{fig:bothspectra}}
\end{figure}

We obtained a total of twelve optical spectra of V4641 Sgr from RCSPEC on the 1.5m SMARTS telescope. RCSPEC is a single-slit grating spectrograph, and gratings used for these data covered approximately 3700~\AA$\!$ to 5400~\AA. Data were taken in 2009 and 2010, over six nights (see Table \ref{table:spectralobs} for details). Spectra were taken in sets of three exposures, at two different times during each night. Each exposure was reduced individually, and the spectrum was extracted and wavelength-calibrated using standard IRAF tasks. Following extraction, each set of three spectra was summed in order to increase the signal-to-noise. These summed spectra were then normalized to their continua. In 2010 spectra were also taken for three comparison stars, with spectral types (B8III, B9III, and A0III) bracketing that which has been proposed for V4641 Sgr (B9III). The data reduction, spectrum extraction, wavelength calibration, and normalization were done in exactly the same manner as for V4641 Sgr. All individual normalized spectra of V4641 Sgr were also shifted to zero velocity by cross-correlating them with a model spectrum for a star with $T_{\mathrm{eff}}=10,500$ K and $\log g=3.5$ \citep{orosz_black_2001} using the IRAF task \texttt{fxcor}. These Doppler-shifted spectra were then summed to produce one composite spectrum (see Figure \ref{fig:bothspectra}), which was used to examine the Balmer absorption lines (see \S \ref{passive:spectra}).

\begin{deluxetable*}{cc}
\tablewidth{0pt}
\tablecaption{Dates of spectral observations, for V4641 Sgr and spectral comparison stars  \label{table:spectralobs}}
\tablehead{\colhead{Date} & \colhead{Objects} }
\startdata
2009 Aug 13 & V4641 Sgr \\
2010 Jun 27 & V4641 Sgr, HIP 88270 (B9III), HIP 93132 (B8III), HIP 105779 (A0III) \\
2010 Jun 28 & V4641 Sgr, HIP 88270, HIP 93132, HIP 105779 \\
2010 Jun 29 & V4641 Sgr, HIP 88270, HIP 93132, HIP 105779 \\
2010 Jun 30 & V4641 Sgr, HIP 88270, HIP 93132, HIP 105779 \\
2010 Jul 2 & V4641 Sgr, HIP 88270, HIP 93132, HIP 105779 
\enddata
\tablecomments{All spectra were taken when V4641 Sgr was in the passive optical state.}
\end{deluxetable*}

Echelle spectra of V4641 Sgr were obtained in service mode using the UVES spectrograph on the second 8.2m VLT telescope at Paranal Observatory (European Southern Observatory), on 2001 June 18, 19, and 22 as part of the ESO program 67.D-0384(A). The slit width was set to 1.2 arcseconds, yielding a resolving power of $R\approx33,000$.  UVES is a double-armed instrument, and we obtained four 2460-second exposures in the standard ``DIC1 346+580'' setting and four 2460-second exposures in the standard ``DIC2 437+860'' setting. By combining these two settings, we obtained essentially complete spectral coverage from 3000 \AA\ to 10,000 \AA. IRAF tasks were used to process these CCD frames to remove the electronic bias, apply the flat-field calibration, and to extract and wavelength calibrate the spectra. The signal-to-noise ratio in the combined Doppler-corrected UVES spectrum was 100 or more per pixel over most of the wavelength range.

\section{Active \& Passive Optical States}\label{activepassivesect}

\citet{Cantrell2008} defined active and passive states for A0620-00 in the following manner: in the passive state all observed variability is consistent with orbital variations and photometric errors; in the active state, the source is typically both brighter and more intrinsically variable than the passive state for an extended period of time. 
If the companion star in a system is bright, the passive state is dominated by the orbital changes of the companion star alone. If the accretion disk contributes a significant fraction of light, the passive state is dominated by the orbital changes of the combined disk/companion star system. These states persist on timescales of weeks or months. 

\begin{figure}[b]
	\epsscale{1.1}
	\plotone{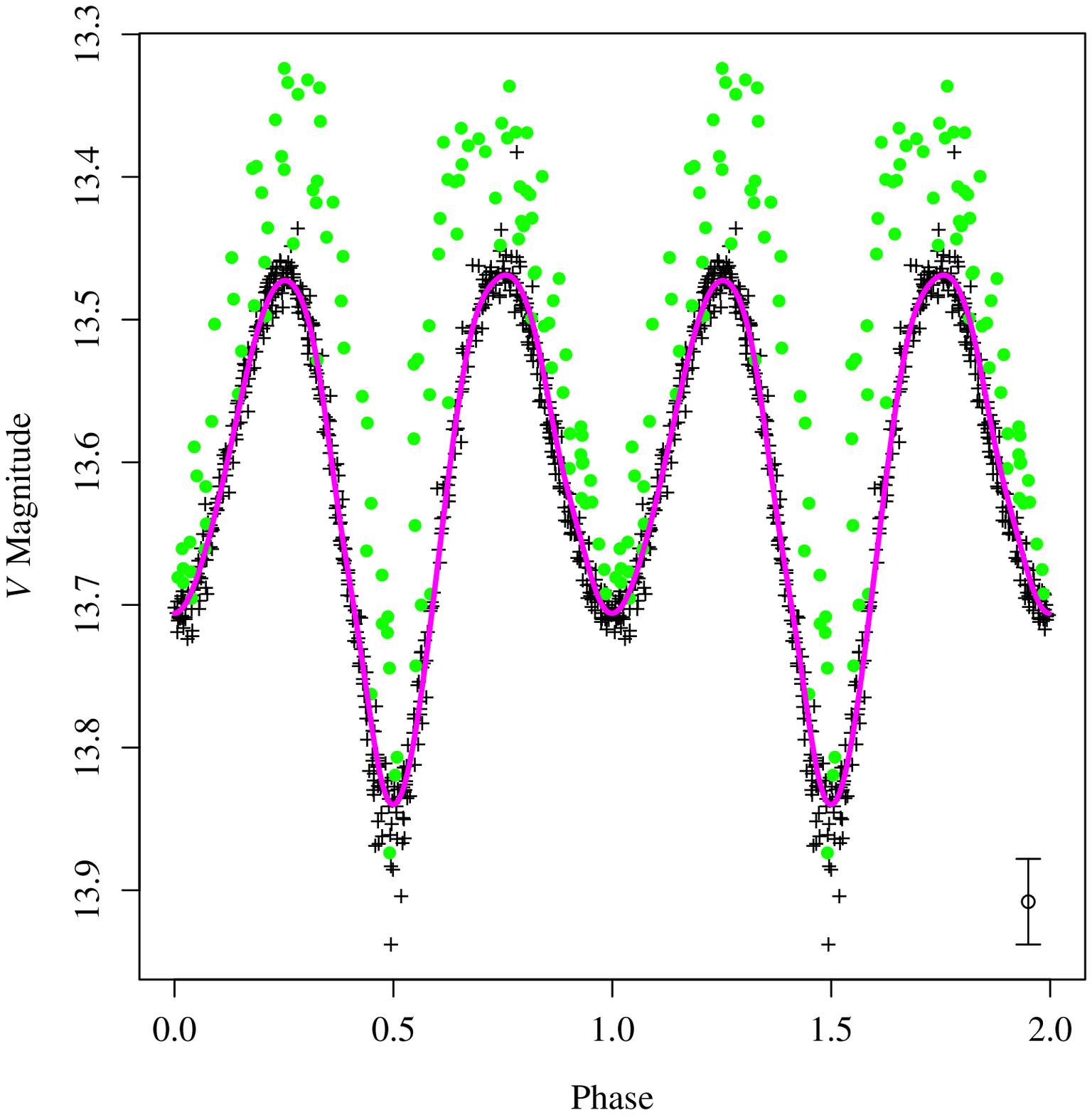}
	\caption{$V$ band V4641 Sgr light curve, folded on orbital period. The entire SMARTS photometric dataset for the $V$ band is included here. After determination of active and passive states as described in the text, the light curve is plotted here to show passive data (black plus symbols), active data (filled circles, green in the online version), and the passive envelope (thick curve, magenta in the online version) which was subtracted from the data in order to analyze the aperiodic variability present in the system. Points are plotted through two orbital periods for clarity. A typical error bar for the data is shown in the lower right corner. \label{fig:v4641-vpassiveenv}}
\end{figure}

In order to determine whether V4641 Sgr has active and passive periods similar to those of A0620-00, the light curve was first folded on the orbital period, which was taken from \citet{orosz_black_2001} (see Figure \ref{fig:v4641-vpassiveenv}). This revealed a clear ``lower envelope'' to the data, which resembled pure ellipsoidal variations. This ellipsoidal variation (or ``passive envelope'') needed to be removed before further analysis of other non-ellipsoidal variability could take place.

Although we used the orbital period given in \citet{orosz_black_2001}, we used a value for $T_{0}$ slightly different from theirs. Specifically, when we folded our photometric data on the \citet{orosz_black_2001} orbital period and $T_{0}$, we found that both maxima and both minima occurred early by 0.03 in phase. We therefore adopted a value of $T_{0} = 2452423.647 \pm 0.01$ as our photometric $T_{0}$, in order to fix this slight inconsistency.

\subsection{Definitions of States}\label{statedefnssect}
The passive envelope was computed for each filter as follows: magnitude data were binned in phase, with 20 equal-width bins total over the range of phase from zero to one. In each bin, the mean and standard deviation of the magnitude points were calculated, all points outside of one standard deviation were discarded, and the mean of the remaining points was calculated. This second, more restrictive, mean was defined to be the location of the passive envelope in that bin. Spline interpolation connected these means together to produce a smooth curve. We note that this method works here because of the large number of passive data points compared to active; in a source with more active points than passive, this method would fail. The results from the $V$ band are shown in Figure \ref{fig:v4641-vpassiveenv}; the passive envelope is shown as a thick curve overlying the data. This smooth curve was subtracted from all the data, thus removing the periodic orbital signature from each light curve. It is important to note that this ``passive envelope'' curve is not a model --- it does not make any assumptions about any physical characteristics of the system, it is an empirical description only.

\begin{figure*}[tb]
	\epsscale{1.1}
	\plottwo{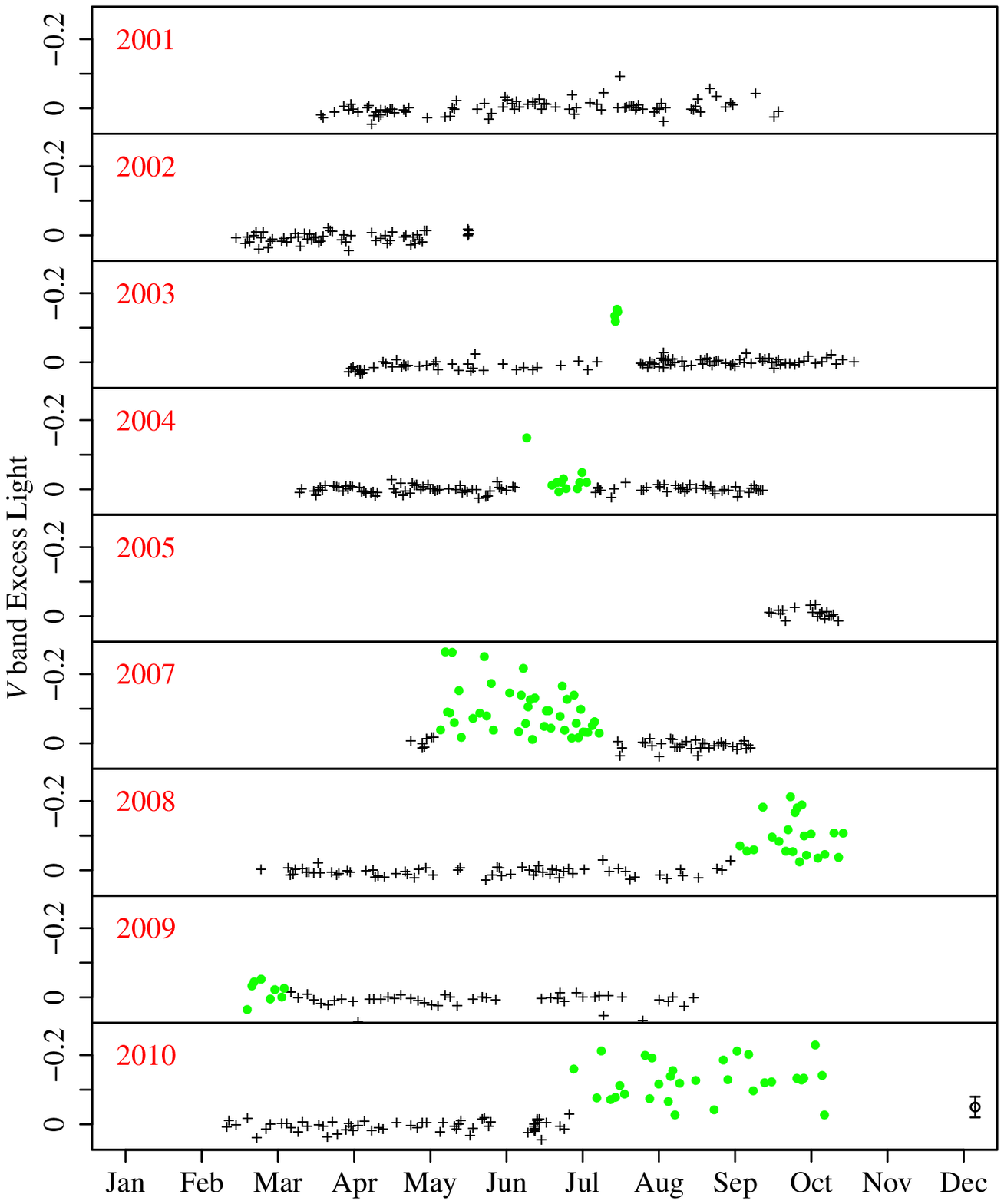}{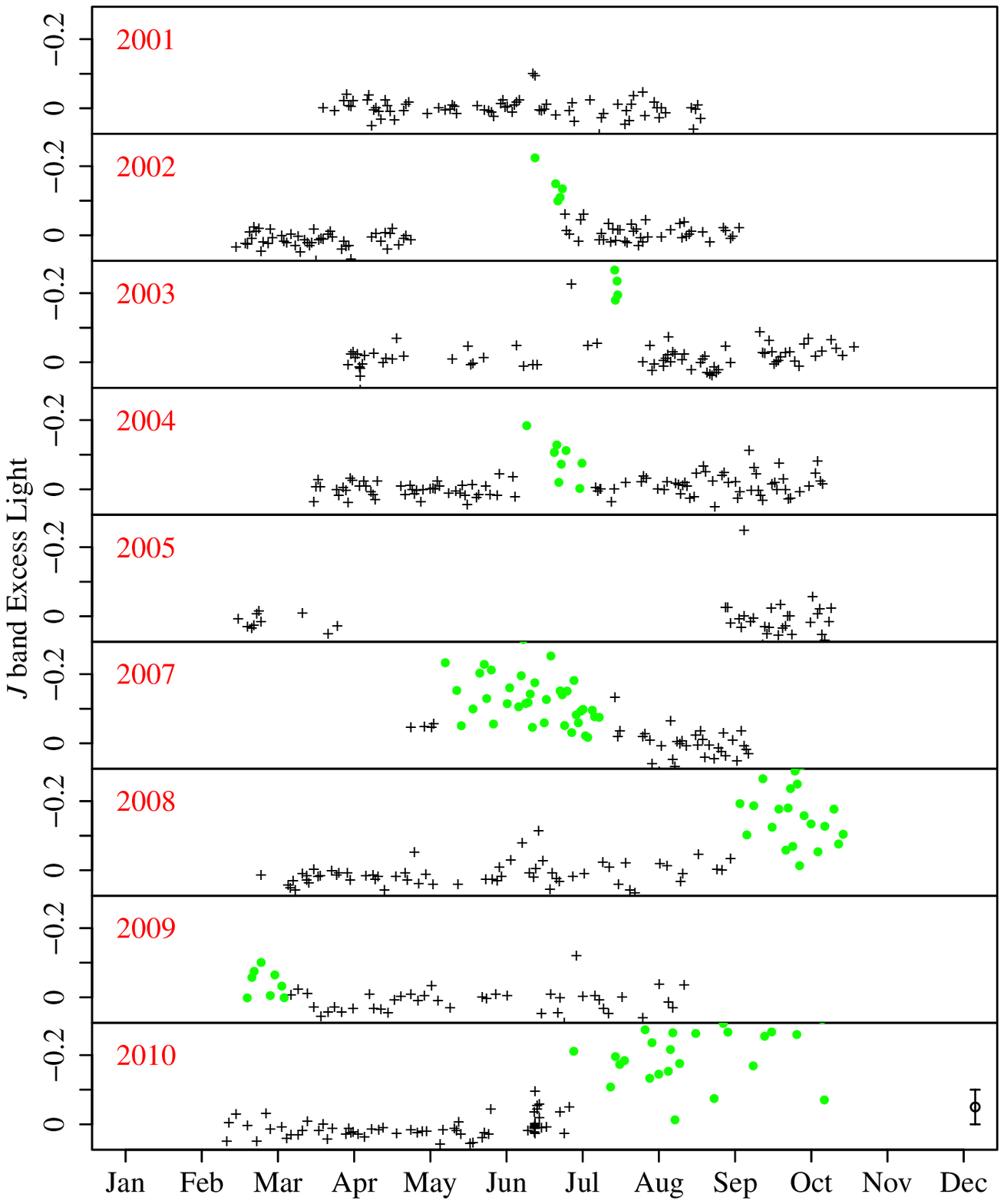}
	\caption{Example light curves for V4641 Sgr, after subtraction of the passive envelope. Left panel shows the $V$ band light curve;   \label{fig:v4641-vbandyearbyyearcolor} right panel shows the $J$ band light curve. \label{fig:v4641-jbandyearbyyear}
	Notice that most of the data fall within a narrow range near zero in both panels ($\pm \sim$0.04 mag, for $V$ band; $\pm \sim$0.1 mag, for $J$ band). Periods of time when most of the data were above this narrow range and when the overall scatter was higher (in at least 2 of the 6 bands), were defined to be ``active''. Passive time periods are marked by black plus symbols, active time periods are marked by filled circles (green in the online version). Typical error bars for the data are shown in the lower right corner of each panel.}
\end{figure*}

We examined the entire data set after the passive envelope was subtracted (as an example, see the $V$ band and $J$ band data in Figure \ref{fig:v4641-vbandyearbyyearcolor}). These light curves clearly showed that most of the data in each band fell in a fairly narrow range clustered around zero. The different optical states were defined by first drawing threshold lines on the plots. The threshold for the optical bands was 0.04 mag, while the threshold for the infrared bands was 0.1 mag. Periods of time when most data were below the threshold were classified as passive for that band, while those times when most data were above the threshold, and where the scatter between points was larger, were classified as active for that band. To avoid spurious classifications, we required that data be classified as active in two or more bands in order to define a particular time period as active. When available, data from all optical bands ($B$, $V$, \& $I$) were examined in order to find active periods. When only IR data were taken, $J$ \& $H$ bands together were used to find active periods (the $K$ band data showed much more intrinsic scatter at all times, and was judged not to be a reliable measure of activity by itself). Final results of this analysis, showing when V4641 Sgr was in each state, are shown in Table \ref{activepassivetimes}.

\begin{deluxetable}{cc}
\tablewidth{0pt}
\tablecaption{Dates of passive \& active states,\\2001 –-- 2010 \label{activepassivetimes}}
\tablehead{\colhead{Dates} & \colhead{State} } 
\startdata
2001 Mar 27 -- 2001 Oct 12 & Passive \\ 
2001 Oct 13 -- 2002 Feb 17 & \nodata \\
2002 Feb 18 -- 2002 Jun 6 & Passive \\
2002 Jun 7 -- 2002 Jun 26 & \nodata \\
2002 Jun 27 -- 2002 Jul 9 & Active \\
2002 Jul 10 -- 2002 Sep 25 & Passive \\
2002 Sep 26 -- 2003 Apr 7 & \nodata \\
2003 Apr 8 -- 2003 Jul 28 & Passive \\
2003 Aug 1 -- 2003 Aug 2 & Active \\
2003 Aug 13 -- 2003 Nov 14 & Passive \\
2003 Nov 15 -- 2004 Mar 16 & \nodata \\
2004 Mar 17 -- 2004 Jun 19 & Passive \\
2004 Jun 20 -- 2004 Jul 19 & Active \\
2004 Jul 24 -- 2004 Oct 31 & Passive \\
2004 Nov 1 -- 2005 Feb 11 & \nodata \\
2005 Feb 12 -- 2005 Nov 6 & Passive \\
2005 Nov 7 -- 2007 May 4 & \nodata \\
2007 May 5 -- 2007 May 15 & Passive \\
2007 May 17 -- 2007 Jul 25 & Active \\
2007 Aug 4 -- 2007 Sep 30 & Passive \\
2007 Oct 1 -- 2008 Feb 28 & \nodata \\
2008 Feb 29 -- 2008 Sep 15 & Passive \\
2008 Sep 24 -- 2008 Nov 8 & Active \\
2008 Nov 10 -- 2009 Feb 21 & \nodata \\
2009 Feb 22 -- 2009 Mar 10 & Active \\
2009 Mar 21 -- 2009 Sep 5 & Passive \\ 
2009 Sep 6 -- 2010 Feb 13 & \nodata \\
2010 Feb 14 -- 2010 Jul 13 & Passive \\
2010 Jul 14 - 2010 Oct 31 & Active
\enddata
\tablecomments{See sect. \ref{activepassivesect} for definitions of the states and details about how data were classified. Lines marked ``\nodata'' represent epochs when we do not have data.}
\end{deluxetable}

\begin{figure*}[tb]
	\epsscale{1.0}
	\plotone{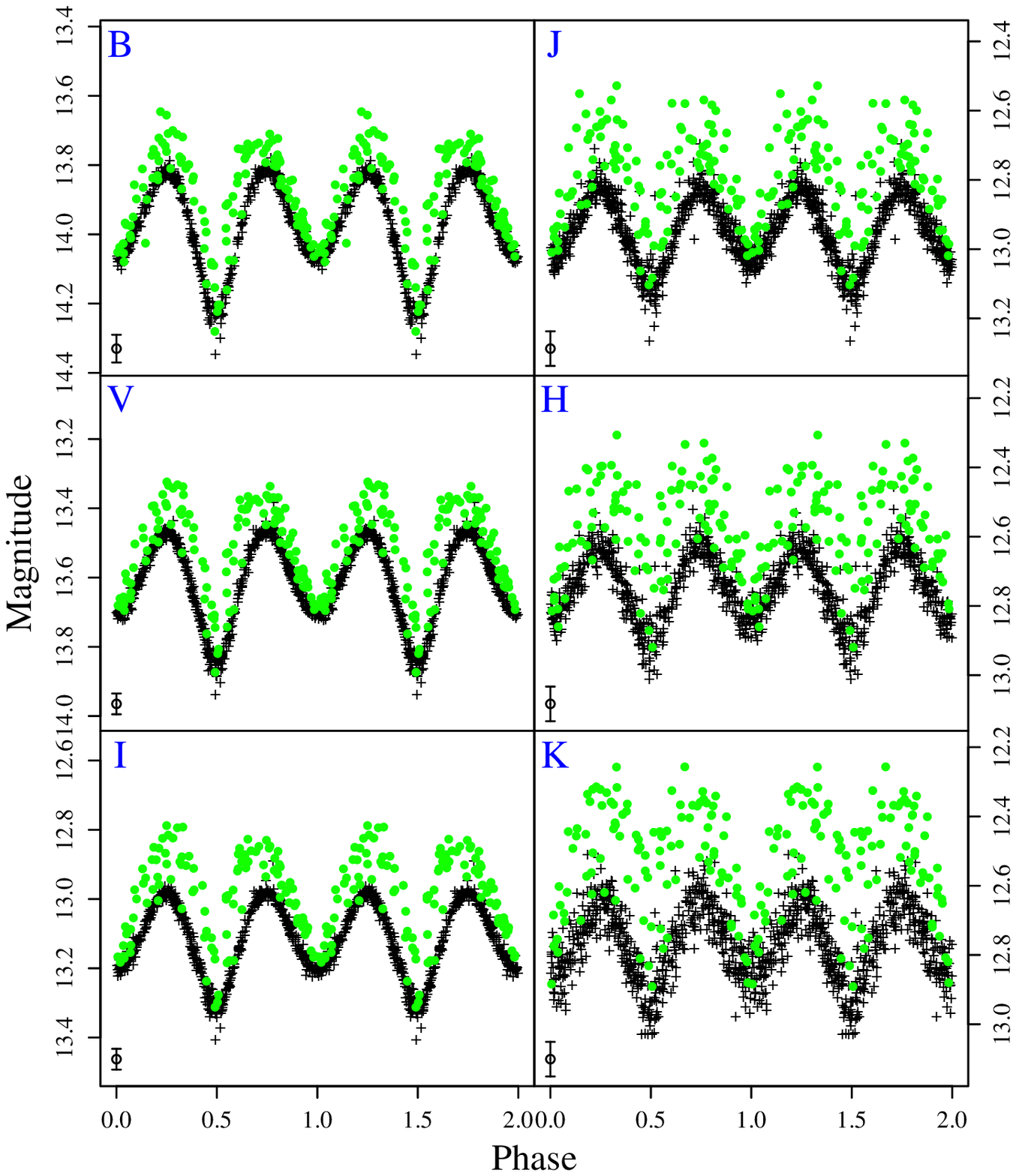}
	\caption{V4641 Sgr light curves in all six optical and infrared filters, folded on the orbital period. The entire SMARTS photometric dataset is included here. The filter is indicated in upper left corner of each panel. Note that vertical axis is \textit{different} for each panel, although the range in magnitudes is the same for all (each vertical axis spans 1.0 mag). Panels are shown so that the secondary minimum is located at approximately the same vertical position in each plot. After determination of active \& passive time periods, different symbols were chosen to mark active and passive states: plus symbols are passive and filled circles (green in the online version) are active.	Points are plotted through two orbital periods for clarity. Typical error bars are shown in the lower left corner of each panel. \label{fig:v4641all6bands}}
\end{figure*}

Light curves in all six bands, folded on orbital period, are shown in Figure \ref{fig:v4641all6bands}. There are significant differences between the bands. In particular, the difference between the primary and secondary minima is largest in $B$ and smallest in $K$. Overall scatter among data is largest in $K$ and smallest in $B$. The difference between the passive data and the brightest active data is largest in $K$ and smallest in $B$.

The optical passive state for V4641 Sgr is remarkably stable. The folded light curve does not change shape over time, unlike A0620-00, which changes shape over timescales of months or years \citep{Cantrell2010}. This contrasting behavior is likely due to the differences in the companion stars in these two systems. V4641 Sgr has a B subgiant as its companion, which is large and brighter than the accretion disk. It therefore takes large changes in behavior of the accretion disk to change the total emission from the system. A0620-00, on the other hand, has a K dwarf as its companion, which is small and cooler than the accretion disk. Changes in that accretion disk can more easily dominate the emission from the system.

Based on the data we examined, during X-ray quiescence V4641 Sgr spends approximately 85\% of the time in the optical passive state and 15\% of the time in the optical active state. It should be noted, however, that we do not have complete time coverage over the span of our data (2001-2010), and additional active and passive states could have been missed.

We constructed color-magnitude diagrams in order to determine if there was an intermediate ``loop'' state in V4641 Sgr, as was found in A0620-00 by \citet{Cantrell2008}. This loop state in A0620-00 was characterized by being intermediate between passive and active in both luminosity and scatter, and by the presence of clockwise loop patterns traced out by successive data points in a color-magnitude diagram. For V4641 Sgr, the relatively long active periods in 2007 and 2010 were examined for this type of behavior. No coherent patterns were found between successive points in a color-magnitude diagram during either active period. Thus V4641 Sgr does not seem to possess an easily identifiable ``loop'' state.

\subsection{The Passive State}\label{colorsedsect}

\subsubsection{Photometric Study}\label{passive:phot}
We compared the passive colors of V4641 Sgr to the colors of a late B star. For the calculations in this section, we used the median magnitude of the passive data near phase 0.0 (secondary minimum) in each band. Secondary minimum was chosen as that is the phase of the orbit when the star is between the observer and the accretion disk. We are therefore looking directly at the back side of the companion star, where it looks most like a normal spherical star.

\begin{figure}[t]
	\epsscale{1.15}
	\plotone{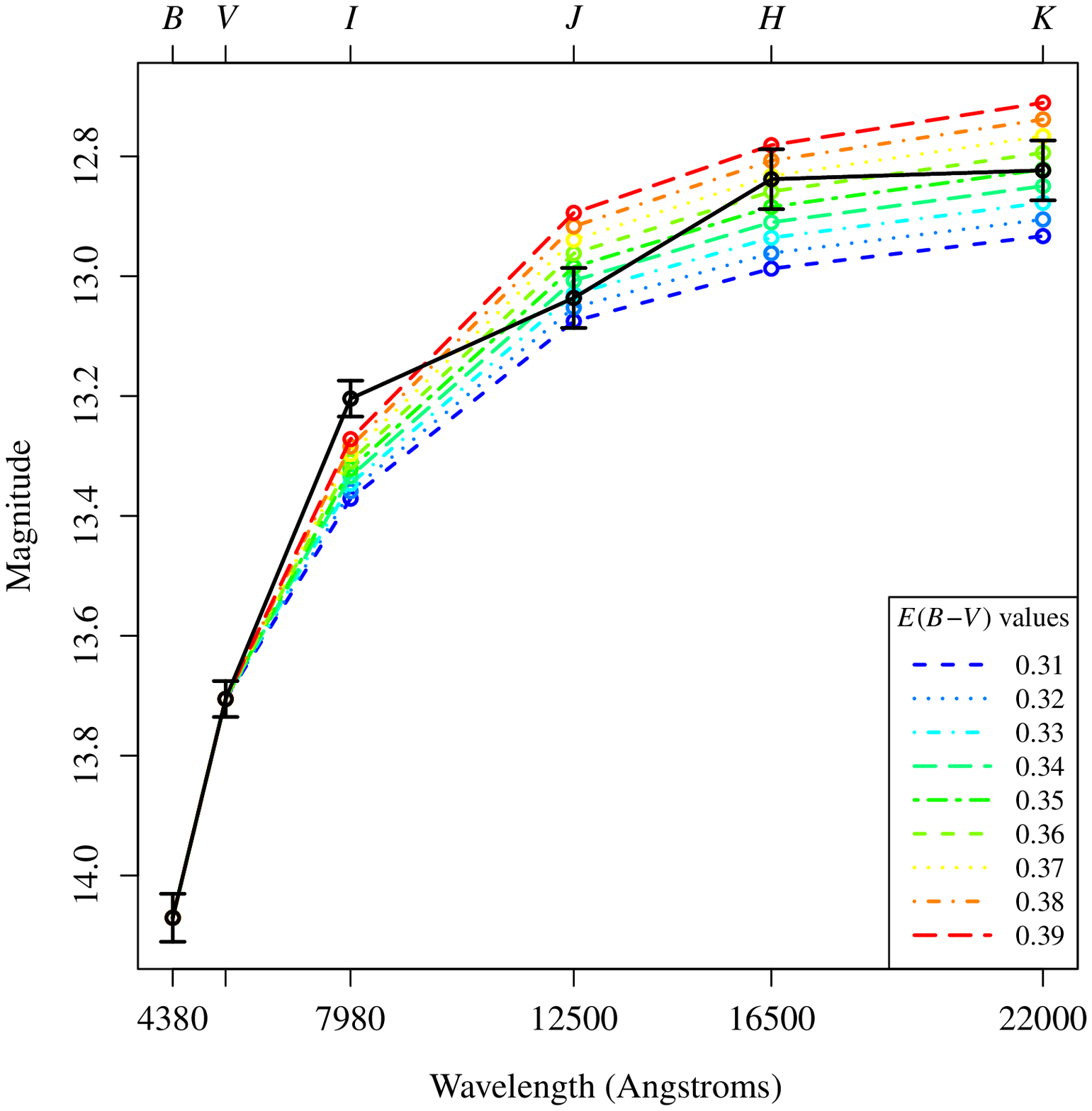}
	\caption{V4641 Sgr spectral energy distribution (SED), compared to reddening a B9III star, with several different possible reddening values. The solid black line is the V4641 Sgr SED; the dotted and dashed lines (shown in different colors in the online version) are the reddened B star SEDs. The reddened B star SED was normalized by matching the \bv color of V4641 Sgr.
	\label{fig:reddeningvalues}}
\end{figure}

To compare our source to a B star, we examined the spectral energy distributions (SEDs) of V4641 Sgr and of a generic reddened B9III star. Using the colors of an unreddened B9III star from \citet{wegner_intrinsic_1994} as a starting point, we reddened the SED of a generic comparison B star, and normalized it by matching its reddened color to the median passive $\bv$ color of V4641 Sgr. Figure \ref{fig:reddeningvalues} shows the SEDs of both V4641 Sgr and the reddened B9III star, with several possible values for the reddening. It is clear that in the passive state V4641 Sgr is consistent with a reddened B star for a range of  $E(\bv)$ values from 0.32 to 0.38.

Recently \citet{gonzalez_reddening_2011} published reddening and metallicity maps of the Galactic Bulge region which have both higher resolution and larger coverage area than previous studies. They have provided an online calculator, the Bulge Extinction And Metallicity (BEAM) Calculator\footnote{\anchor{http://mill.astro.puc.cl/BEAM/calculator.php}{http://mill.astro.puc.cl/BEAM/calculator.php}}, that allows one to retrieve the reddening value for any part of their mapped region. This online calculator gives a value of $E(J\!-\!K) = 0.1927 \pm 0.1005$ for V4641 Sgr, which converts to $E(\bv) = 0.37 \pm 0.19$ using the relations of \citet{cardelli_relationship_1989}. This is consistent with the values we found by reddening a generic B star, and we adopt this value for the reddening along the line of sight to V4641 Sgr. 

The reddening along the line of sight to this source has previously been quoted differently in several sources in the literature: \citet{orosz_black_2001} use $E(\bv)$ = 0.32 $\pm$ 0.1, \citet{int_zand_bepposax_2000} use 0.24 $\pm$ 0.1, \citet{chaty_optical_2003} use $\approx 0.25$, and \citet{lindstrm_new_2005} find $E(\bv)$ could be as high as $0.4 - 0.6$. Given the relatively large errors, these values are all broadly consistent with that from the BEAM calculator.

\begin{figure}[b]
	\epsscale{1.15}
	\plotone{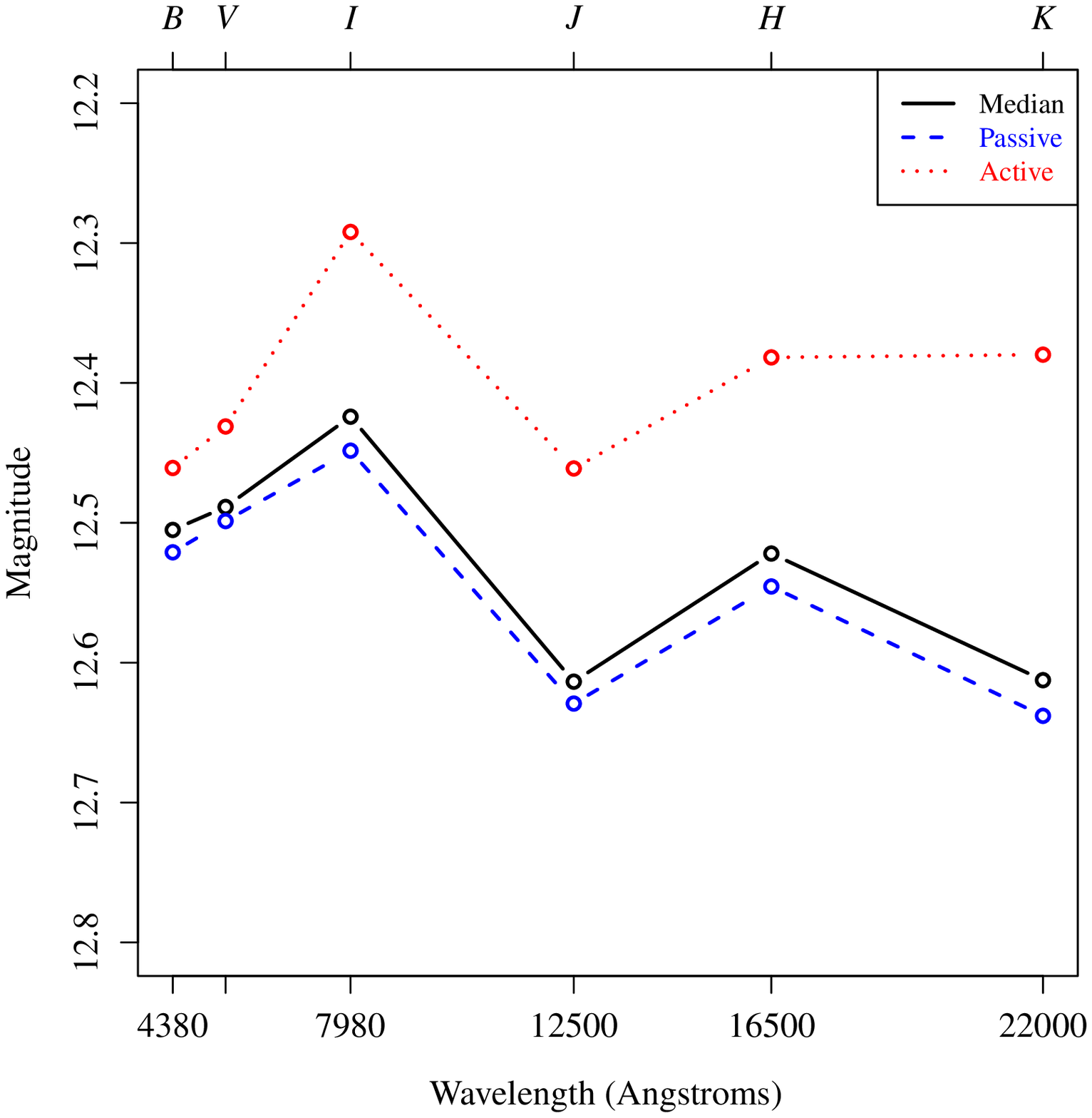}
	\caption{Dereddened spectral energy distributions of V4641 Sgr, showing the differences between the optical states. The solid line labeled ``Median'' is the median of all data (passive and active together). The dashed line (blue in the online version) is passive data only, and the dotted line (red in the online version) is active data only. Note that the active state is always brighter than the passive, and that this difference increases toward the red. Dereddening errors are dominated by the error in $E(\bv)$; final errors on these dereddened magnitudes are 0.79 mag in $B$, 0.60 mag in $V$, 0.36 mag in $I$, 0.18 mag in $J$, 0.12 mag in $H$, and 0.08 mag in $K$. \label{fig:sed}}
\end{figure}

We calculated extinction coefficients in the other bands using $A_{\lambda}/A_{V} = a(x) + b(x)/R_{V}$ and the relations for $a(x)$ and $b(x)$ given in \citet{odonnell_rnu-dependent_1994} for the optical bands and in \citet{cardelli_relationship_1989} for the infrared bands. 
The errors on the dereddened absolute magnitudes of V4641 Sgr are dominated by the error in $E(\bv)$, and come to 0.79 mag in $B$, 0.60 mag in $V$, 0.36 mag in $I$, 0.18 mag in $J$, 0.12 mag in $H$, and 0.08 mag in $K$.
The final dereddened SED for V4641 Sgr is shown in Figure \ref{fig:sed}, and highlights the differences between the median total (passive \& active together) SED, the passive-only SED and the active-only SED.

\subsubsection{Spectroscopic study}\label{passive:spectra}
We analyzed the SMARTS optical spectra of V4641 Sgr to determine the fraction of light coming from the accretion disk in this system. All of our spectra were taken when V4641 Sgr was in the passive optical state. Because the companion star in this system is very bright, we expected the spectroscopic data to be dominated by the companion star but to possibly contain a small, essentially constant, amount of disk light. This is quite different from other X-ray binary systems, which can have significant contributions ($30-60\%$) from the accretion disk at all times \citep{Cantrell2010,filippenko_mass_1995}.

We calculated the fraction of the light due to the accretion disk in V4641 Sgr by analyzing the equivalent widths of the Balmer absorption lines. B stars are known to have deep Balmer absorption lines, as our source does. Accretion disks have continuum emission and also often double-peaked emission lines, either of which can dilute the absorption features from the companion star. Measuring the equivalent width of the Balmer lines in V4641 Sgr and comparing to the equivalent width of the same lines in the comparison B star, therefore, allowed us to determine whether there was any emission from the accretion disk contaminating these spectral features. 

After examining the spectra of both V4641 Sgr and our comparison B9III star HIP 88270 (see Figure \ref{fig:bothspectra}), we found that our spectra allowed the possibility of a small contribution from the disk (up to 15\% at a 2$\sigma$ level), but were basically consistent with little to no emission from the accretion disk. This is plausible during the passive state, given the large intrinsic luminosity of the companion star in this system and the fact that the colors are consistent with those of a solitary late B star.

\subsection{The Active State}\label{activelightsect}

\begin{figure*}[tb]
	\epsscale{1.0}
	\plotone{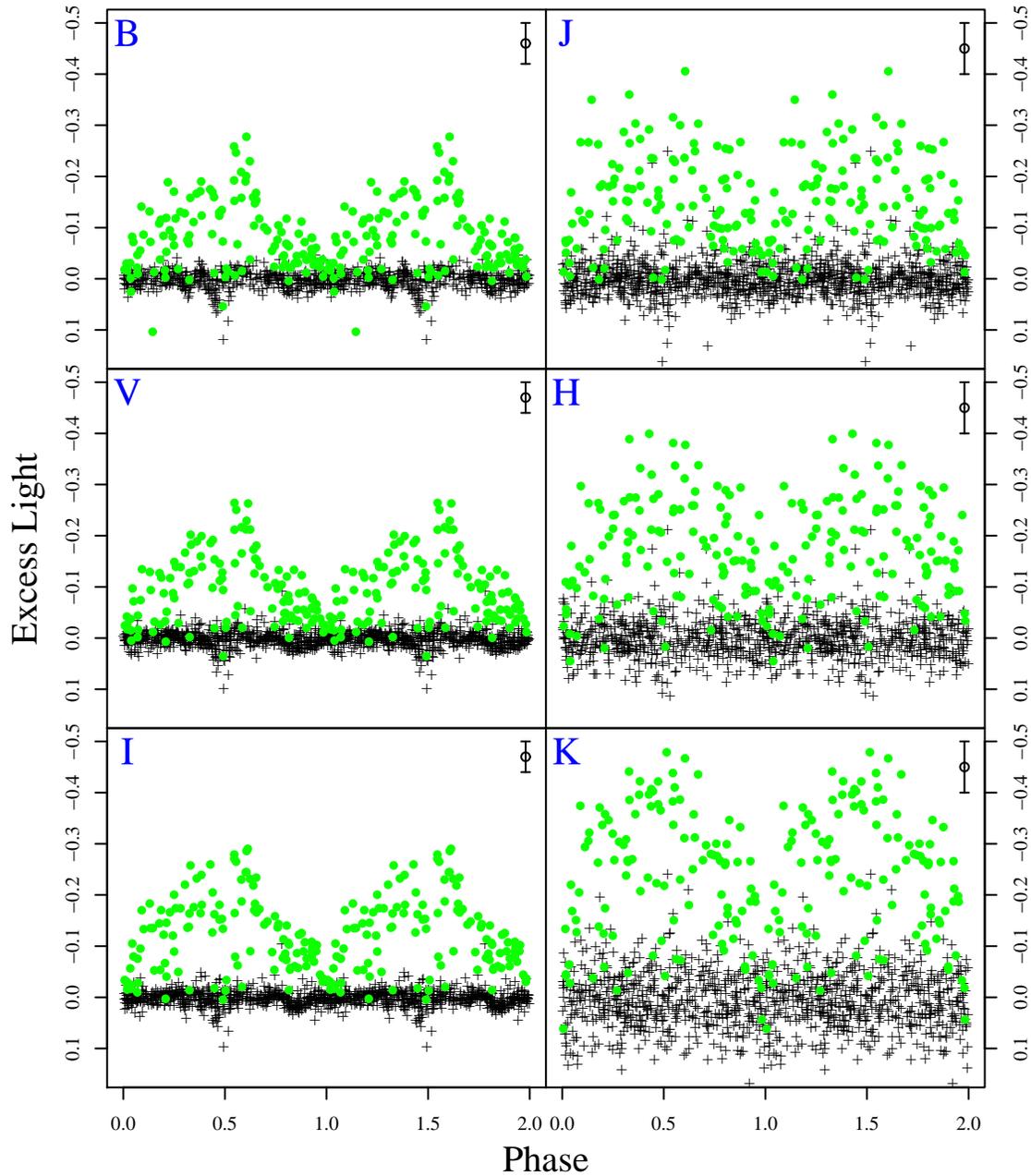}
	\caption{V4641 Sgr light curves as in Figure \ref{fig:v4641all6bands}, except that here the passive envelope has been subtracted in each filter. The entire SMARTS photometric dataset is included here. The filter is indicated in the upper left corner of each panel. Note that the vertical axis here is \textit{the same} for each panel. Plus symbols are passive data and filled circles (green in the online version) are active data. Points are plotted through two orbital periods for clarity. Typical error bars are shown in the upper right corner of each panel. Notice the change in activity throughout the orbital period --- activity is highest near phase 0.5 (when the accretion disk is between the observer and the companion star), and lowest near phase 0.0 (when the companion star is between the observer and the accretion disk). 	
	The small cluster of passive points at phase 0.5 which are fainter than all other data in each band may indicate infrequent slight eclipses of the companion star by the accretion disk. See text for more discussion of these features of the data.
\label{fig:all6bandsexcessonly}}
\end{figure*}

Once optical states for V4641 Sgr were defined and the passive envelope was subtracted from each data point, we examined the resulting residuals, hereafter designated ``excess light''. Plotting the excess light folded on the orbital period (see Figure \ref{fig:all6bandsexcessonly}) revealed that the active state excess light was not consistent throughout the orbital period. Rather, the active light was brighter near phase 0.5 (primary minimum), and fainter near phase 0.0 (secondary minimum). This increase in brightness at primary minimum is likely related to changes in the viewing angle of the system throughout the binary orbit. Primary minimum is the point in the orbit when the accretion disk is between the observer and the companion star. At this phase, we could be seeing more emission from processes originating in the accretion disk, such as irradiation of the face of the companion or an accretion disk wind. Given that the system is at a moderately high inclination angle (see \S \ref{finalparamvalues}), these types of emission might be partially blocked during the rest of the orbital period by the rather large companion star.

The small cluster of passive points at phase 0.5 which are fainter than all other data in each band are puzzling. They are too faint to be explained by ellipsoidal variations --- attempting to fit those faintest points while calculating the inclination angle of the system leads to angles $>90\degr$. They could be explained by an infrequent slight eclipse of the companion star by the relatively dark accretion disk, however we do not have enough of these very faint data to draw firm conclusions.

\begin{figure*}[t]
	\epsscale{1.15}
	\plottwo{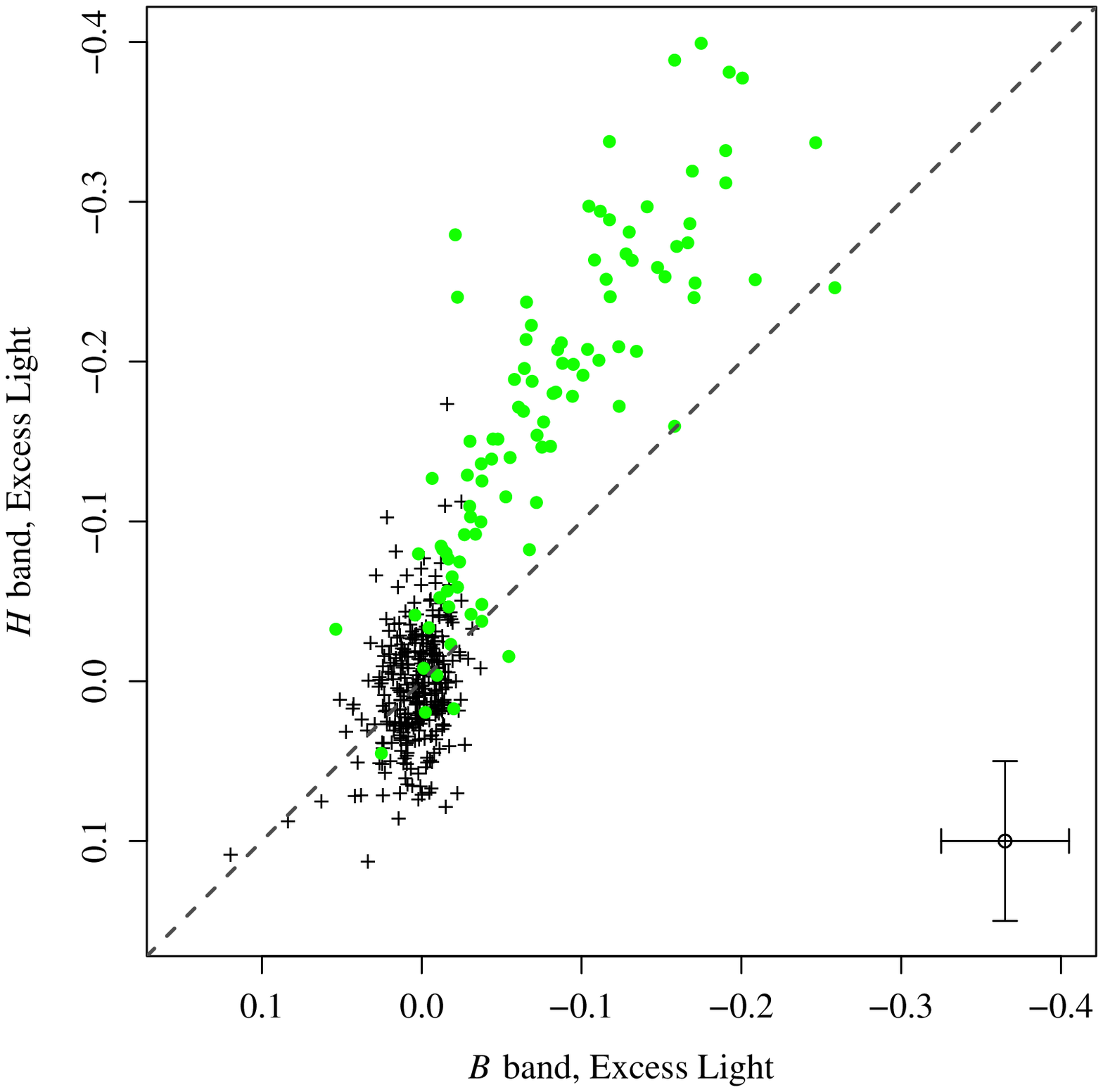}{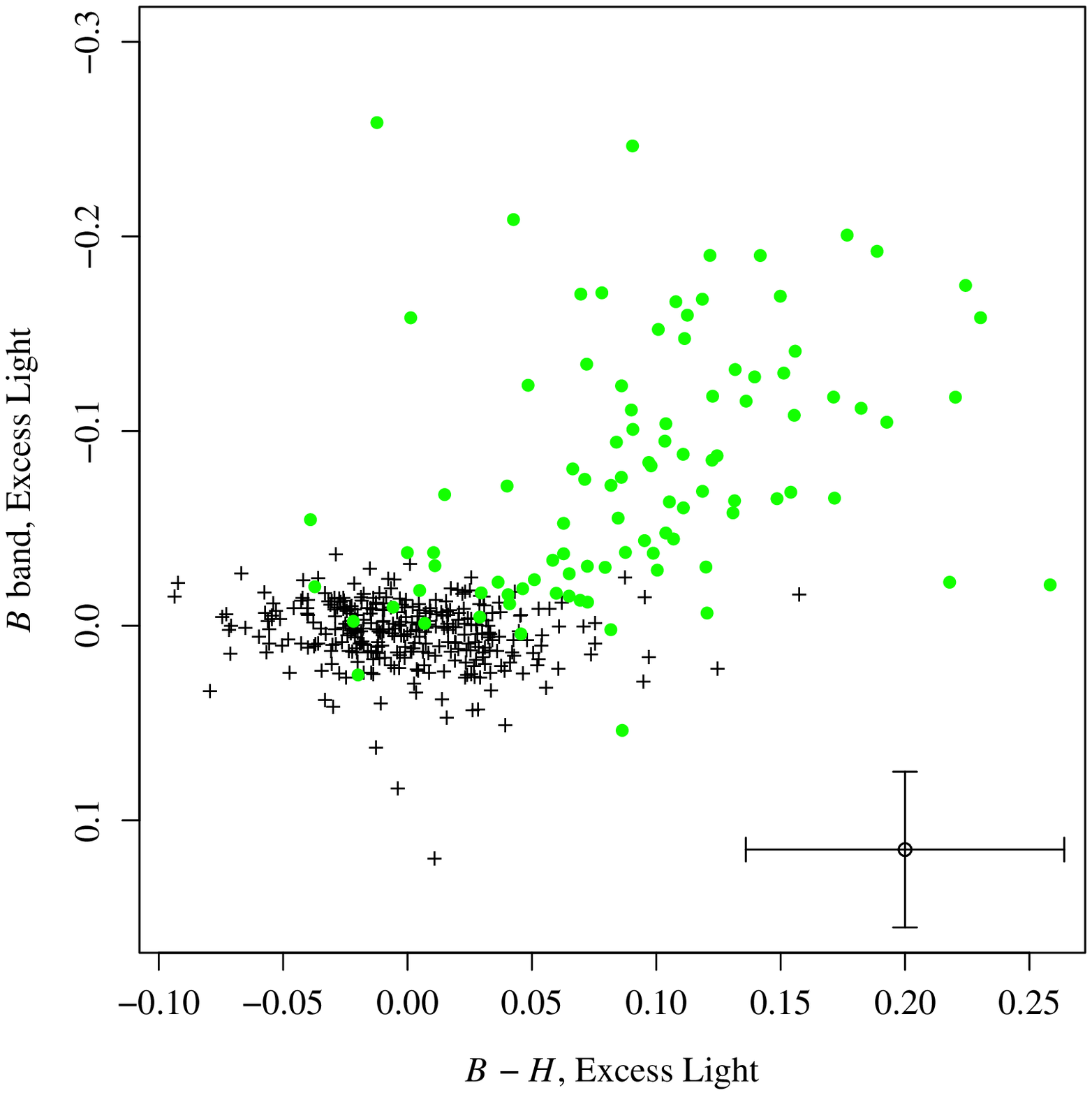}
	\caption{Both panels show excess light only. Plus symbols are passive data, filled circles (green in the online version) are active data. Typical error bars for data are shown in the lower right corner of each panel.
	\textit{Left panel}: Excess light in bands $H$ \& $B$ plotted against each other. The dashed line indicates where a one-to-one relation would lie. The majority of the active data lie above this line, indicating that the source brightens less in $B$ than in $H$ during an active state --- that is, the source gets redder than the passive data when it gets brighter in the active state. 	\label{fig:deltahdeltab} 
	\textit{Right panel}: Color-magnitude diagram showing $B$ vs. $B\!-\!H$ in excess light. This shows, as in the left panel, that in the active state the source gets redder than the passive data when it gets brighter. 
	\textit{Both}: These panels demonstrate that the passive data represent emission dominated by the blue companion star, while the active data represent heightened emission from the accretion flow in the system (which is hotter than the passive accretion flow, but still cooler than the companion star).
	\label{fig:colormagbvsbminush}}
\end{figure*}

We investigated emission from the accretion flow in V4641 Sgr by examining the colors of the excess light during active states. We selected $B\!-\!H$, to explore the largest separation between blue and red possible with these data. Although $K$ band is obviously redder than $H$, $K$ band data displayed much more intrinsic scatter at all times, making it more difficult to draw firm conclusions using that band. Figure \ref{fig:deltahdeltab} shows the results of this investigation. In the left panel we plot excess light in the $H$ band against excess light in the $B$ band, while in the right panel we plot $B$ magnitude of the excess light against $B\!-\!H$ color. 
The active data represent heightened emission from the accretion flow in the system (which, although hotter than the passive accretion flow, is still cooler than the companion star), while the passive data represent emission dominated by the blue companion star. Therefore the excess light emission is redder during an active state than it is during a passive state. 
This contrasts with A0620-00, where the active light is bluer than its companion, a K star \citep{Cantrell2008}. 
This raises the possibility that the accretion flow in the two systems exists at similar temperatures and produces similar emission. Differences in color between V4641 Sgr and A0620-00 could simply be attributed to their very different companion stars.

\begin{figure*}[b]
	\epsscale{1.15}
\plottwo{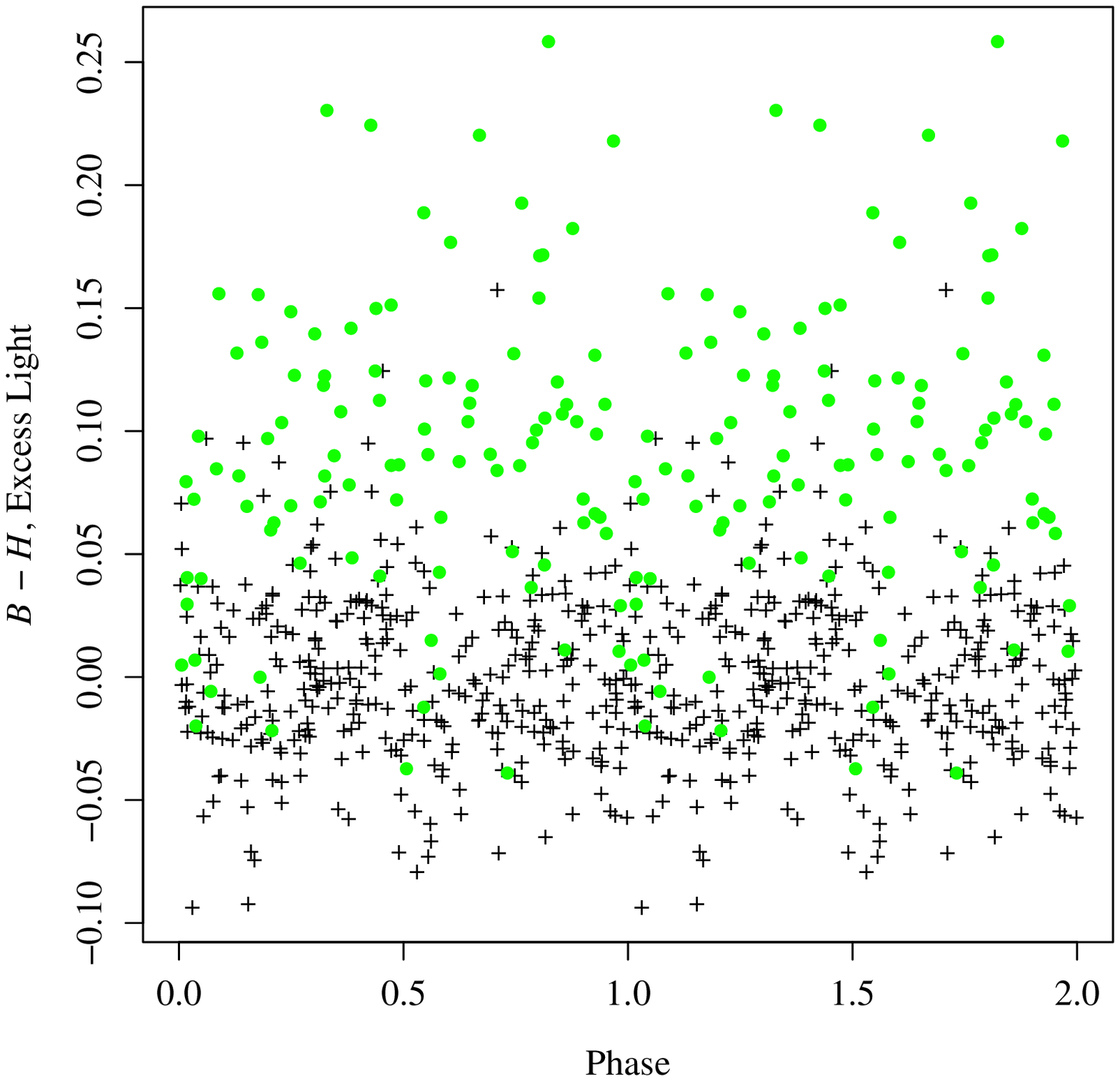}{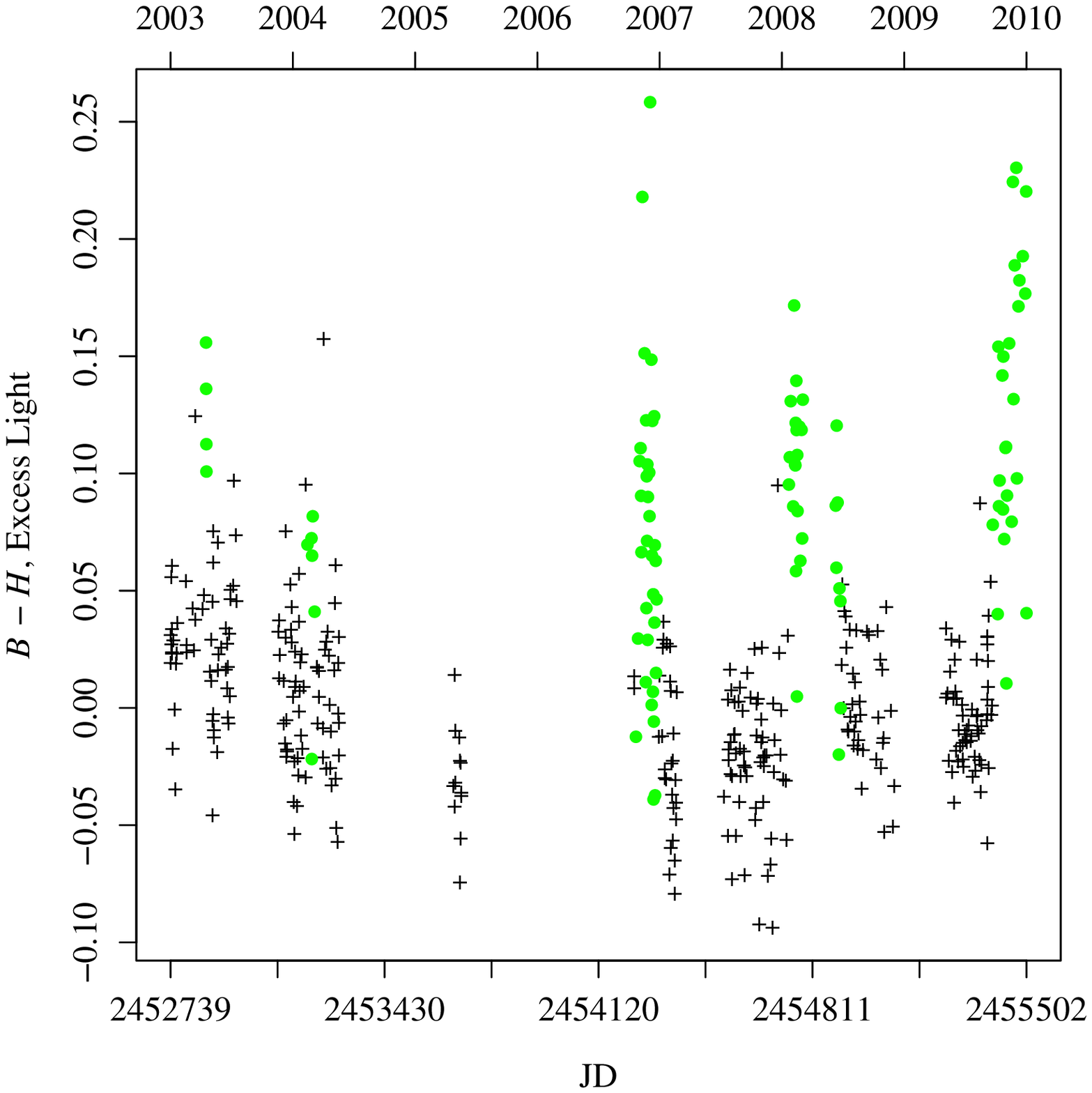}
	\caption{$B\!-\!H$, excess light only, as described in \S \ref{activelightsect}. Data toward the top of both panels are redder, data toward the bottom are bluer. Vertical axes are the same in both panels. Plus symbols are passive data, filled circles (green in the online version) are active data.
	\textit{Left panel}: $B\!-\!H$ color folded on orbital phase. Color during active periods does not seem to change much over the course of the orbit. Points are plotted through two orbital periods for clarity. \label{fig:bminushvsphase}
	\textit{Right panel}: $B\!-\!H$ color shown over time. Note the changes between states (rapidly changing from red to blue and back). 	\label{fig:colorvstime}}
\end{figure*}

In order to explore potential secular changes in the accretion disk of V4641 Sgr, we examined the $B\!-\!H$ color of the excess light over time. The $B\!-\!H$ color of the excess light, plotted against orbital phase (see left panel of Figure \ref{fig:bminushvsphase}), does not seem to show obvious trends over the course of the orbit. This suggests that changes in the accretion disk during an orbit affect the optical emission and the infrared emission equally, rather than causing large changes in one with respect to the other. Plotting the $B\!-\!H$ color against time (see the right panel of Figure \ref{fig:colorvstime}), does not seem to show any trend over the course of our ten years of observations.

\subsubsection{Excess Light Spectral Energy Distributions}

We converted our photometric measurements to flux (in Jy) in order to examine the excess light in more detail, specifically to investigate the SED of the excess light in different optical states. V4641 Sgr calibrated magnitudes and passive envelope values for all bands were converted to flux units using flux zeropoints $B_{0}=4063$ Jy, $V_{0}=3636$ Jy, $I_{0}=2416$ Jy \citep{bessell_model_1998}, $J_{0}=1670$ Jy, $H_{0}=980$ Jy, and $K_{0}=620$ Jy \citep{frogel_photometric_1978,elias_infrared_1982}. After this conversion, the passive envelope was subtracted from the total flux to give excess light measurements in Jy. We note that this conversion and subtraction process was actually performed twice, once on the original calibrated magnitudes and once on the dereddened magnitudes. This yielded two sets of excess light flux measurements, one as observed and one dereddened.

\begin{figure*}[t]
	\epsscale{1.15}
	\plottwo{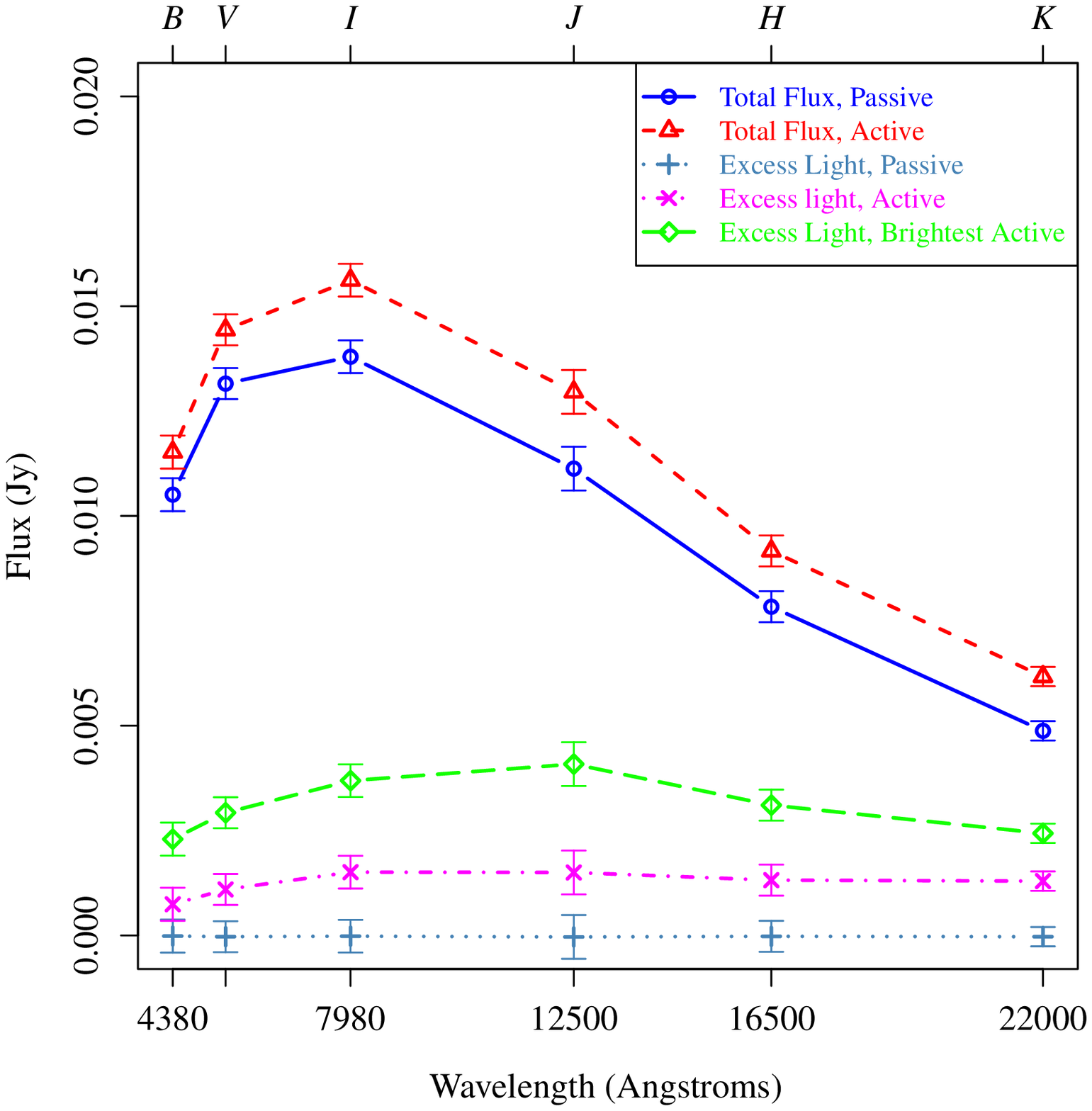}{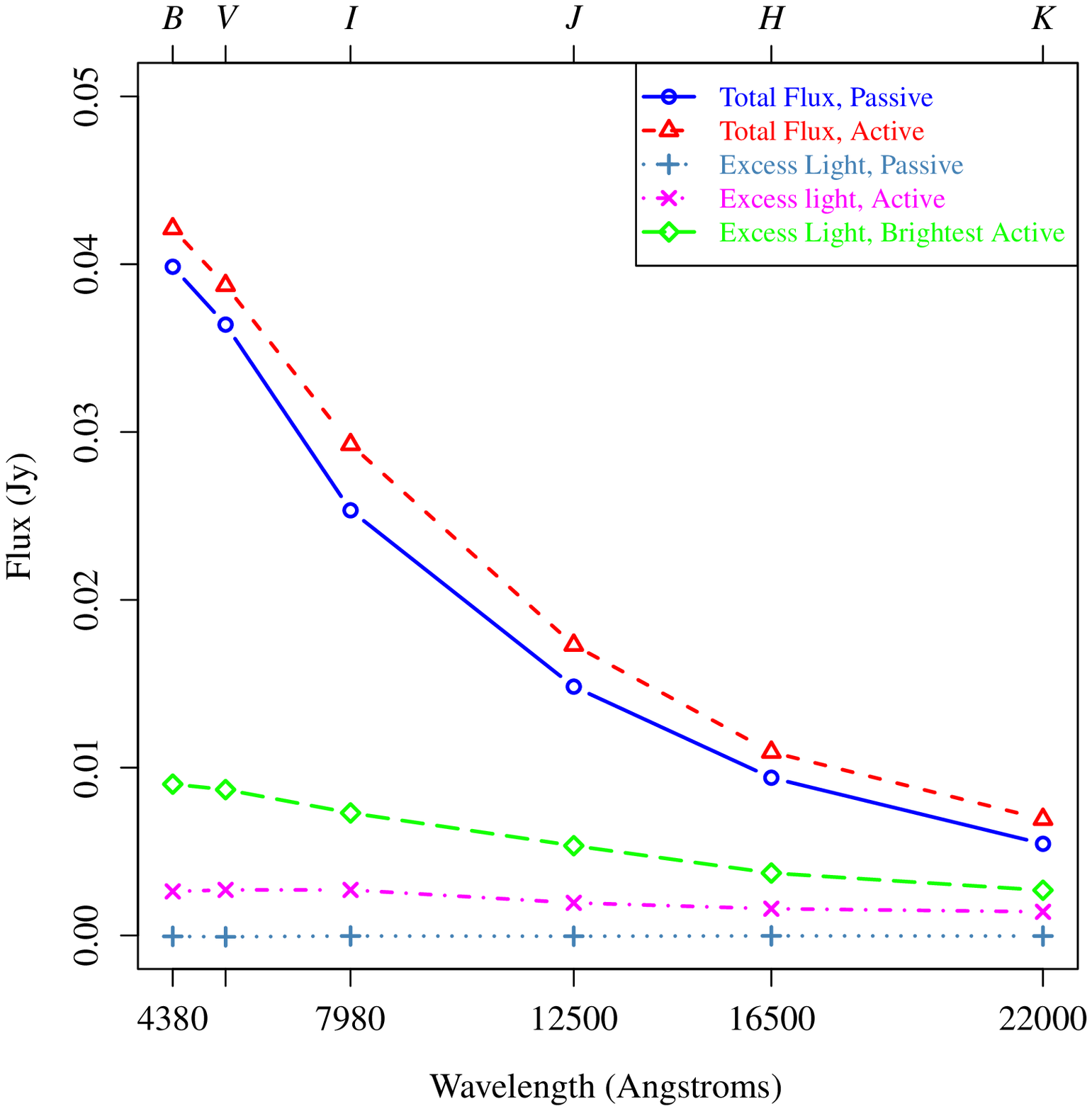}
	\caption{SEDs of V4641 Sgr in flux units (Jy). The right panel is the dereddened version of the left panel.
	For both panels, the upper two lines (dashed line with triangles and solid line with open circles) show the total flux, before subtraction of the passive envelope, while the lower three lines (long-dashed, dot-dashed, and dotted) show the excess light only. (Color version is online: red dashed line, blue solid line, green long-dashed line, magenta dot-dashed line, and light-blue dotted line.)
	Notice that the active data are brighter than the passive data in both the total flux and in the excess light. 
	 Dereddening errors for these data are dominated by the error in the reddening measurement, and are 0.029 Jy in $B$, 0.020 Jy in $V$, 0.009 Jy in $I$, 0.002 Jy in $J$, and 0.001 Jy in both $H$ and $K$.
	 \label{fig:sedsflux}}
\end{figure*}

Figure \ref{fig:sedsflux} shows a variety of spectral energy distributions for V4641 Sgr in flux units. The right panel is the dereddened version of the left panel. For both panels, the upper two lines show the total flux (before subtraction of the passive envelope), while the lower three lines show the excess light only. Both the total flux and the excess light have been separated into active data and passive data (taking the median of the data in each state), and for the excess light we have also separated out the brightest 5\% of the active data.

Several features can be observed in Figure \ref{fig:sedsflux}. In both the total flux and in the excess light only, the active state data are brighter than the passive state; this agrees with what is seen in the folded light curves in Figure \ref{fig:v4641all6bands}. In examining the excess light only, we see that the passive state is near zero, indicating that there is little emission in this state that does not come from the companion star. To compare the different active state data, we calculated the ratio of the brightest 5\% of active excess light (diamond symbols) to the median active excess light (``x'' symbols): $F_{\mathrm{brightest}} / F_{\mathrm{active}}$ = 3.09 for $B$, 2.67 for $V$, 2.45 for $I$, 2.72 for $J$, 2.36 for $H$, and 1.88 for $K$ (these values represent the data as observed, however using dereddened data gives similar ratios). The ratios are largest at the blue end of this spectral range, and smallest at the red end, indicating that the brightest 5\% of active excess light are bluer than the median active excess light. This agrees with the idea that the active optical state is associated with a heightened, and therefore hotter, accretion flow in the disk.

\section{Determination of Component Masses}\label{elcsect}

\subsection{Overview of Ellipsoidal Light Curve Modeling}\label{elcdetails}

The Eclipsing Light Curve (ELC) code \citep{orosz_use_2000} was used to fit ellipsoidal models to the observed passive light curves in order to determine geometric and radiative parameters of V4641 Sgr. ELC generates models of ellipsoidal light curves, with possible contributions from the disk, star spots, and X-ray heating of the companion. Eclipses of both the star and the disk are included, and can be relevant at high inclination angles. For a given star and disk configuration, it computes the light curve which would be observed through a filter with a user-specified transmission curve. It has several optimizer routines to search the wide parameter space of possible models, finding optimal fits to data based on $\chi^2$ minimization. Our method has been described by \citet{Cantrell2010}. Here we give a brief description of the relevant parameters in ELC, while \S\ref{v4641inputs} details our specific use of ELC to fit the light curves of V4641 Sgr.

ELC has many parameters to set the geometric and radiative properties of a binary system. Geometric parameters include the orbital period $P$, the orbital separation $a$, the ratio of the masses $Q = M_{\mathrm{BH}}/M_{2}$, where $M_{\mathrm{BH}}$ is the mass of the black hole and $M_{2}$ is the mass of the companion star, the inclination $i$, and the Roche lobe filling factor $f_{2}$. Radiative parameters include the companion star's average temperature $T_{2}$ and its gravity darkening exponent $\beta$, as well as the choice of either specific intensities derived from model atmospheres or one of several parameterized limb-darkening laws.

The accretion disk in ELC is assumed to be circular and flared. The parameters describing it include the outer radius of the disk $r_{\mathrm{out}}$, the inner radius of the disk $r_{\mathrm{in}}$, the opening angle of the flared outer disk rim $\beta_{\mathrm{rim}}$, the temperature of the inner edge $T_{\mathrm{disk}}$, and the power-law exponent on the disk temperature profile $\xi$, where $T (r) = T_{\mathrm{disk}}(r/r_{\mathrm{in}})^\xi$. 

Free parameters are fit using one of several possible optimization routines, including one based on a genetic algorithm and one using a Monte Carlo Markov Chain. In addition, the same optimizing algorithm can be used several times with the free parameters listed in a different order in the input files, which results in a different initial population for the genetic algorithm or a different random walk for the Markov Chain. 

The optimization routines can produce hundreds or thousands of models, each with a different set of free parameters, a full set of calculated output values, and an associated $\chi^2$ value. A $\chi^2$ value is calculated for each light curve (that is, for each filter for which one has provided data), and for each observed quantity which is input, using the uncertainties included with the photometry and the observables. These are summed to give the final value for that model. The model with the lowest total $\chi^2$ value is given as the final, best-fitting model.

Note that we use the term ``free parameter'' to describe a quantity which can be optimized in order to generate the best-fitting model; we use ``physical quantity'' or ``output quantity'' to describe a value which is calculated from the model once the free parameters have been optimized. 

Some clarification about observable quantities as inputs: stellar or system properties which have been measured previously (such as mass, radius, surface gravity, effective temperature, rotational velocity, radial velocity, duration of an X-ray eclipse, inclination angle, mass ratio, or disk fraction) can be given to ELC as additional model constraints. These properties are not, however, used to freeze the values of free parameters which would otherwise be optimized, or to force ELC to only explore values within the given uncertainties. Rather, each model gives predictions for all possible output quantities, whether or not input values have been provided, based on its particular combination of free parameters. These predictions are then compared to input observed values (and their given errors), and a $\chi^2$ is computed. These $\chi^2$ values are calculated for each input measured quantity and are then added to the total $\chi^2$ value assigned to a model. 

It is worth noting that it is possible for ELC to produce an output value which is completely inconsistent with the given input value. Thus an additional way to evaluate the best-fitting model is to explicitly compare all input observables with their output predicted counterparts. If the predictions closely match the measured values, then the model is more likely to be correct -- that is, more likely to be a reasonable representation of the physical properties of the binary system.

\subsection{Modeling V4641 Sgr}\label{v4641inputs}

We used the already well-determined value of 2.81730 days for the orbital period \citep{orosz_black_2001}.
We assumed the companion star fills its Roche lobe, and consequently we set $f_{2} = 1$. We also assumed the star rotates synchronously and that the orbit is circular.

The value of $T_{2}$ for V4641 Sgr has been spectroscopically determined to be $T_2=10,500\pm 200$ K \citep{orosz_black_2001,sadakane}, and based on this temperature the gravity darkening exponent was set to $0.25$ \citep{orosz_black_2001,claret_studies_2000}. The effects of X-ray heating were neglected, as there is effectively no X-ray flux in quiescence ($L_{X} \ll L_{*}$). The specific intensities used for V4641 Sgr were derived from the NextGen grid \citep{hauschildt_nextgen_1999a,hauschildt_nextgen_1999b} with updates.

The inner radius of the disk $r_{\mathrm{in}}$ was fixed at $0.001$ times the black hole's Roche lobe radius. Since optical emission is dominated by the outer disk, the precise value of the inner radius is unimportant to the optical emission discussed here. 

To set a temperature for the accretion disk, we examined our data folded on the orbital period, and found that the amplitude of the $V$ band light curve from the maximum to the primary minimum (at phase = 0.5) is $\sim$0.3 mag. This amplitude is primarily determined by the mass ratio ($Q$) and inclination ($i$), however it can be affected by the presence of an accretion disk. If the accretion disk eclipses the star during primary minimum, the amplitude of the observed variation will increase. For this to occur, the disk has to be cooler than the star (cool enough to contribute no light to the system). We therefore constrained the temperature of the disk to be smaller than that of the companion star (a typical value was 5000 K). This is consistent with spectroscopic evidence that the disk does not contribute significantly to the observed optical and near-IR emission in V4641 Sgr. 

In an X-ray binary, it is usually assumed that the source of the X-rays is physically very small compared to the mass-donor companion star. If the X-ray source is eclipsed, the duration of the X-ray eclipse $\Theta$ depends on the inclination, mass ratio, and the Roche lobe filling factor.  Thus a measurement of the eclipse duration provides a strong constraint on the system geometry \citep[for example][]{m33,rawls}. Likewise, a lack of observed eclipses sets an upper limit on the inclination for a given mass ratio \citep{chanan}. 

In the case of V4641 Sgr, the X-ray outburst flares are bright, but short-lived, so that while bright flux can show the lack of an eclipse, low flux during a flaring state cannot show the existence of one. Furthermore the bright flares from V4641 Sgr show evidence of super-Eddington outflows and somewhat extended emission  \citep{revnivtsev_super-eddington_2002, maitra_x-ray_2006}, so that the flux might be affected by the size of the source. The assumption should be valid for the lower fluxes of quiescence, and \citet{revnivtsev_super-eddington_2002} found marginal indication that the \textit{RXTE} ASM data prior to the 1999 outburst showed a binary orbital dependence which was consistent with obscuration that could be caused by an inclination of $65-70\degr$.

\begin{figure*}[tb]
	\epsscale{1.1}
	\plottwo{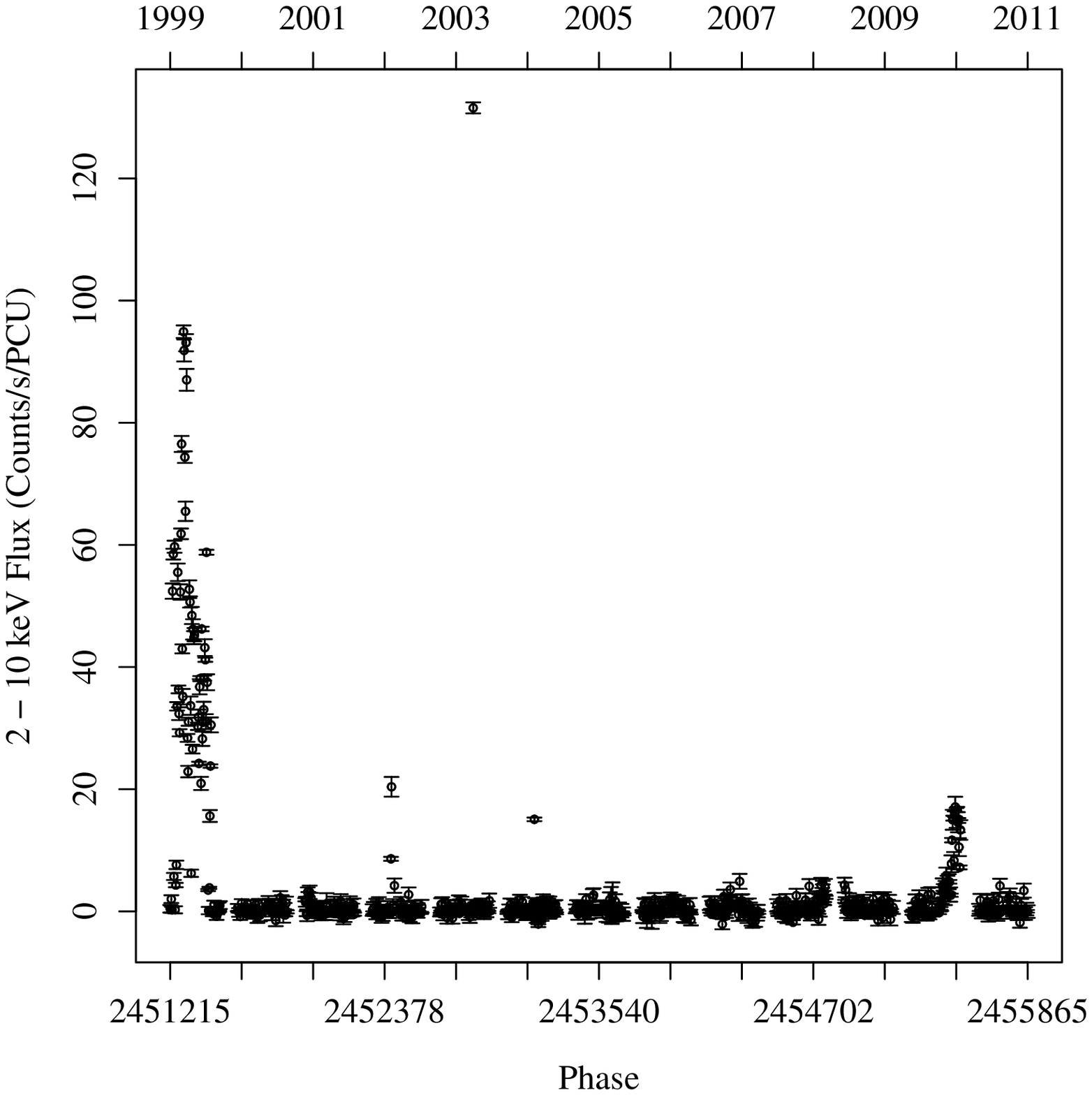}{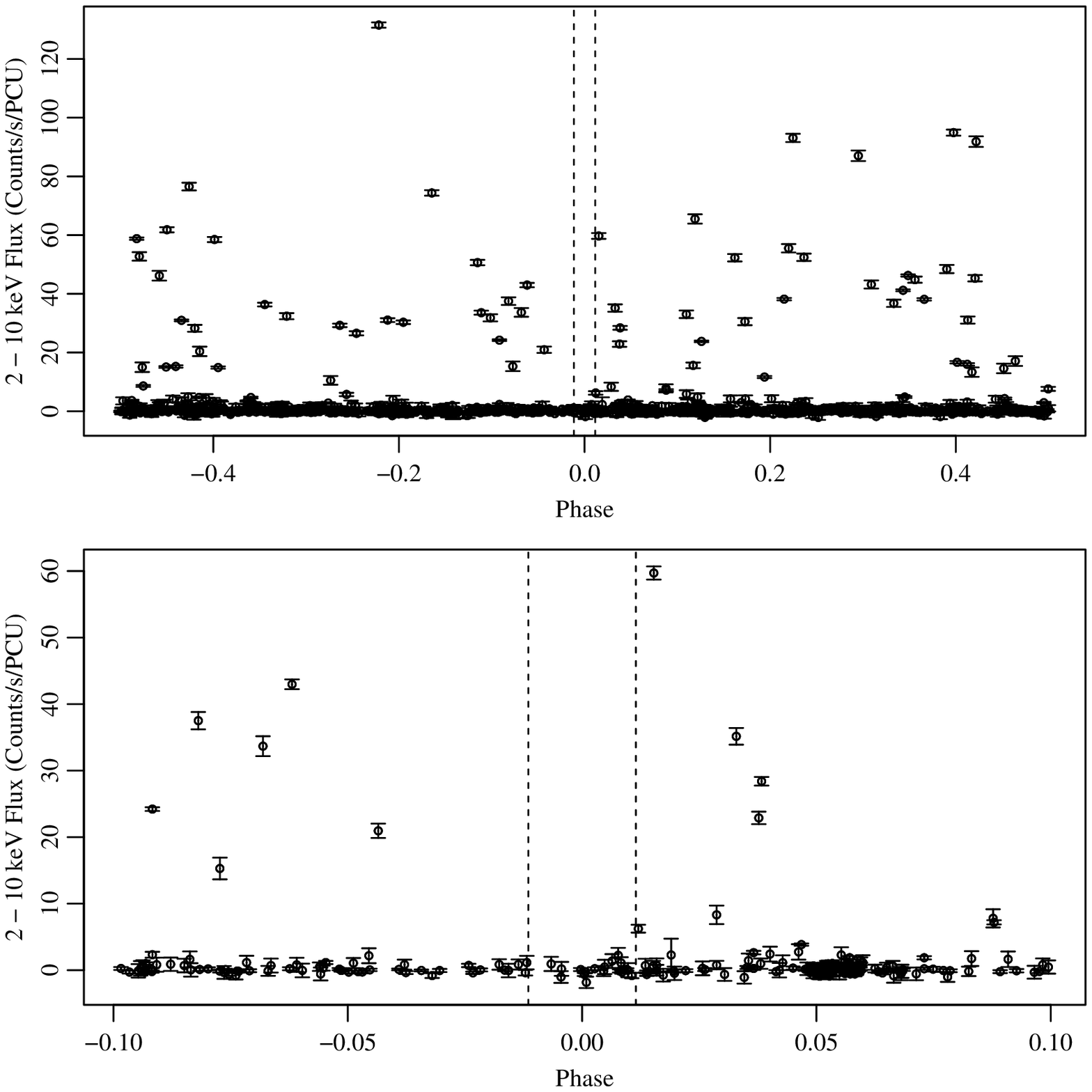}
	\caption{X-ray data taken of V4641 Sgr by the Proportional Counter Array (PCA) on \textit{RXTE}. 
	\textit{Left panel}: Data over time, showing the quiescent flux as well as low-level activity. 
	These data do not include data from flaring outbursts. 
	\textit{Right panel, top}: Data folded on the orbital period.
	\textit{Right panel, bottom}: Same data as top right, zoomed in to show portion of the orbit near phase 0.0, when an eclipse would occur. These data show that if an X-ray eclipse does occur in this system, the maximum duration (indicated by vertical dashed lines) is $8.3 \pm 0.6$ degrees in phase, or $\sim$1.6 hr.
  \label{fig:xrays}}
\end{figure*}

The majority of the data in Figure \ref {fig:xrays} are from scans with the Proportional Counter Array (PCA) on \textit{RXTE} \citep{2001ASPC..251...94S}. Some relatively high points indicate activity in transition to and from flaring outbursts. The gaps are the periods when mission constraints prevented \textit{RXTE} from carrying out the scans. The light curve is supplemented by pointed observations in 2004 which were in the quiescent period of $V$ and $J$ observations as shown in Figure \ref{fig:v4641-vbandyearbyyearcolor}. Pointed observations in 2002, 2003, and 2005 were not during the X-ray quiescent period. One pointed observation in 2010 was during the active part of a quiescent period as identified in $V$ and $J$, but was not included, as the X–rays were at the same level as the 2005 flaring. They were not near phase 0.0, and would not have appeared in the lower right of Figure \ref {fig:xrays}. Thus the latter represents the best PCA constraint on an eclipse duration during X-ray quiescence, and gives an upper limit of $\Theta < 8.3 \pm 0.6{\degr}$.

We ran the ELC optimizing routines with six free parameters: $i$, $Q$, $a$, $r_{\mathrm{out}}$, $\beta_{\mathrm{rim}}$, and $T_2$. 
For each of these we provided a reasonable range of values for ELC to explore: 
 $50^{\circ} < i < 90^{\circ}$, as we know based on previous work that the system is not at a low inclination; 
 $1.5 < Q < 2.4$, encompassing values allowed by current and previous spectroscopy; 
 14 R$_{\sun} < a <$ 22 R$_{\sun}$, based on the size of the companion star; 
 $0.1 < r_{\mathrm{out}} < 0.999$ (expressed as a fraction of the Roche lobe radius), allowing any size for the accretion disk; 
 $0.0\degr < \beta_{\mathrm{rim}} < 2.0\degr$, typical values for accretion disks in such systems; 
 and 
 10,250 K $< T_{2} <$ 10,750 K, bracketing the known temperature of the companion star (we included this particular free parameter to investigate whether modeling would agree with observations). 
We used passive state data in all six bands as the input data to be fit. 

We also provided four observed quantities as external constraints on the modeling process. Three of these observables were a surface gravity of $\log g=3.5\pm 0.1$ dex \citep{orosz_black_2001}, a $K$-velocity of the companion star of $K_2=211.3\pm 1.0$ \citep{lindstrm_new_2005}, and an upper limit on the duration of an X-ray eclipse of $\Theta < 8.3 \pm 0.6^{\circ}$. The final observable was the rotational velocity, which was determined from a combination of spectroscopy and modeling. This is discussed in detail in the next sections. Results from investigating line profiles in the spectra is in the following section, while the modeling efforts are detailed in \S\ref{rotvelmodeling} and \S\ref{sect:vrotsini}. 

All input parameters are summarized in Table \ref{table:inputelcparams}.

\begin{deluxetable}{cc}
\tablewidth{0pt}
\tablecaption{Input parameters to ELC modeling for V4641 Sgr\label{table:inputelcparams}}
\tablehead{\colhead{Parameter} & \colhead{Value / Range} }
\startdata
\multicolumn{2}{c}{\textit{Known Values}} \\
$P$ & 2.81730 d \\
$T_{0}$ & 2452423.647 \\
$f_{2}$ & 1 \\
$\beta$\tablenotemark{(a)} & 0.25 \\
$r_{\mathrm{in}}$ & 0.001\tablenotemark{(b)} \\
$T_{\mathrm{disk}}$ & 5000 K \\
\multicolumn{2}{c}{\textit{Observed Quantities}} \\
$\log(g)$ & $3.5 \pm 0.1$ \\
$K_{2}$ & $211.3 \pm 1.0$ km s$^{-1}$ \\
$\Theta$\tablenotemark{(c)} & $<8.3 \pm 0.6 \degr$ \\
\multicolumn{2}{c}{\textit{Free parameters}} \\
$i$ & $50 - 90 \degr$ \\
$Q$ & $1.5 - 2.4$ \\
$a$ & $14 - 22$ R$_{\sun}$ \\
$r_{\mathrm{out}}$ & $0.1 - 0.999$\tablenotemark{(b)} \\
$\beta_{\mathrm{rim}}$ & $0.0 - 2.0 \degr$ \\
$T_{2}$ & $10250 - 10750$ K
\enddata
\tablenotetext{(a)}{Gravity darkening exponent.}
\tablenotetext{(b)}{Given as a fraction of the Roche lobe radius of the black hole.}
\tablenotetext{(c)}{Duration of X-ray eclipse.}
\end{deluxetable}

\subsubsection{Rotational Velocity from Spectra}\label{rotvelspectra}

A measurement of the projected rotational velocity yields a measurement of the mass ratio $Q\equiv M_{\mathrm{BH}}/M_2$, where $M_{\mathrm{BH}}$ is the black hole mass and $R_L$ is the Roche lobe radius:
$$
{V_{\mathrm{rot}}\over V_2}={V_{\mathrm{rot}}\sin i\over K_2}=
{R_L\over a}\left({M_{\mathrm{BH}} + M_2\over M_{\mathrm{BH}}}\right)
={R_L\over a}\left(1+\frac{1}{Q}\right).
$$

The UVES spectra containing the Mg \textsc{II} $\lambda 4481.2$ line were normalized to the local continuum using low-order spline fits, Doppler-shifted to zero velocity using the updated ephemeris, and averaged to produce a mean Mg \textsc{II} line profile.  Hubeny's IDL program SYNPLOT was used to compute model line profiles, using a Kurucz model \citep{Kurucz} with $T_{\rm eff}= 10\,500$~K and $\log g=3.5$ \citep{orosz_black_2001, sadakane}.

The Mg \textsc{II} $\lambda 4481.2$ line is commonly used to measure the projected rotational velocity of late B and early A stars, since it is relatively free of pressure broadening and line blending \citep{Gray}. The line profile depends mainly on the rotational velocity $V_{\mathrm{rot}}\sin i$ and on the magnesium abundance, with the abundance setting the depth and the rotational velocity setting the line width and the shapes of the line wings. We therefore computed a grid of model line profiles using various values of $V_{\mathrm{rot}} \sin i$ and Mg abundances and compared them to the observed profile using a $\chi^2$ test.  The best-fitting model was for $V_{\mathrm{rot}}\sin i=101.7\pm 0.8$ km s$^{-1}$ and a magnesium abundance of 5.2 times solar. \citet{orosz_black_2001} previously found $V_{\mathrm{rot}}\sin i=124\pm 4$ km s$^{-1}$ and a magnesium abundance of about 7 solar, based on the analysis of a moderate resolution spectrum ($R=7000$).  The differences between their results and ours probably can be attributed to the differences in the spectral resolving power \citep[see][]{steeghs}. \citet{sadakane}, who had spectra with $R=40,000$, found $V_{\mathrm{rot}}\sin i=95\pm 10$ km s$^{-1}$ and a magnesium abundance near solar.  While their rotational velocity is consistent with what we find here, the magnesium abundances are somewhat different.  Our main purpose here is the measurement of the rotational velocity and not the elemental abundances.  A proper abundance analysis would require better base model atmospheres with enhanced abundances built in (we used solar metallicity Kurucz models with abundances tweaked by SYNSPEC) and would account for possible NLTE effects.  Fortunately, the rotational velocity is not correlated with the abundance, as found in studies that measure the rotational velocity from template matching \citep{steeghs}.  

\begin{figure*}[tb]
	\centering
	\includegraphics[angle=-90,scale=0.7]{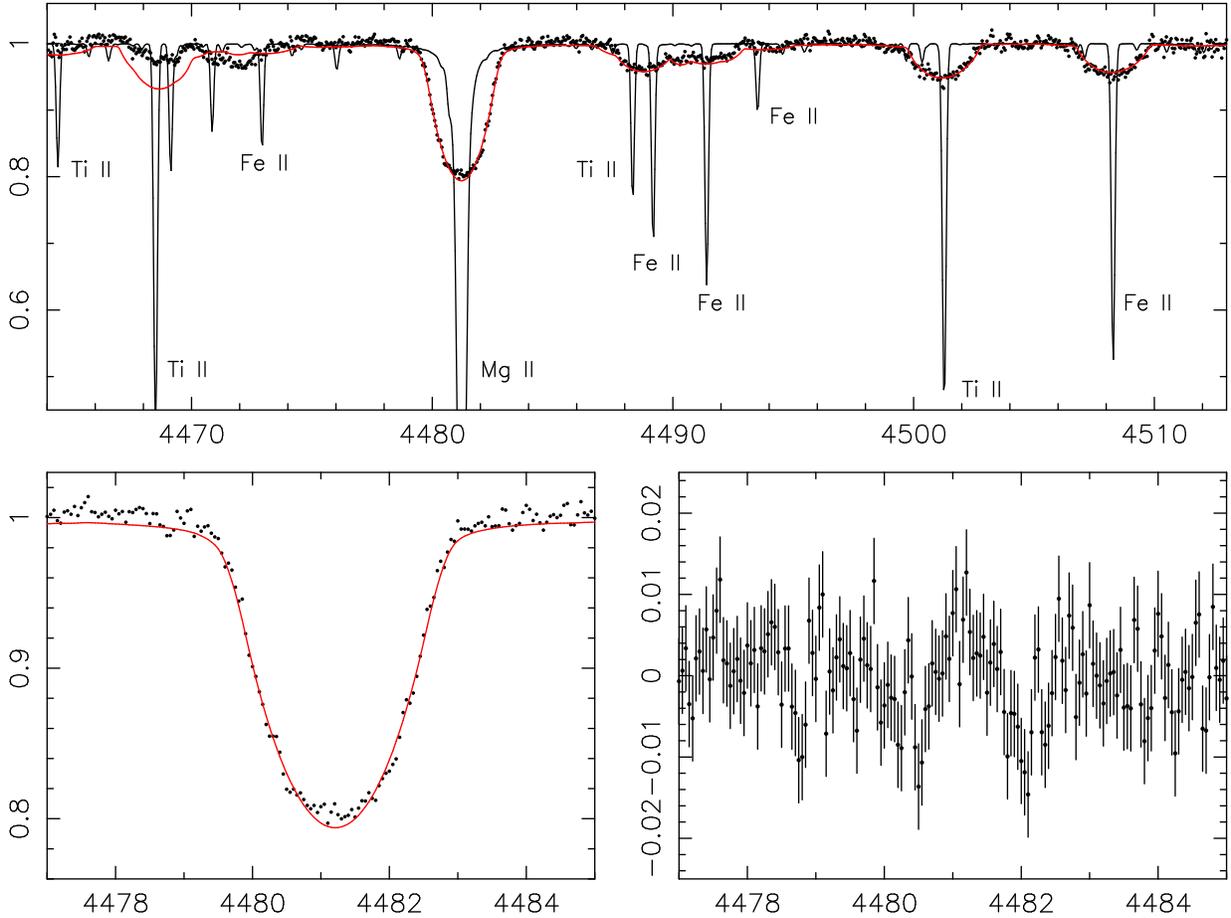}
	\caption{
	\textit{Top}: The normalized UVES spectrum in the region near the Mg \textsc{II} $\lambda 4481.2$ (filled circles) is shown with the best fitting model that has a rotational velocity of $V_{\rm rot}\sin i=101.7\pm 0.8 $ km s$^{-1}$ (solid line; red in the online version). A model with no rotational broadening is also shown as a solid line, with absorption line identifications provided for the stronger features. 
	\textit{Bottom left}:  A close-up view of the Mg  \textsc{II} line (filled circles) with the best-fitting model spectrum.
	\textit{Bottom right}:  The residuals of the model fit to the Mg  \textsc{II} line are shown.
  \label{fig:mgb}}
\end{figure*}

We also computed a grid of models that included several Ti \textsc{II} and Fe \textsc{II} lines between 4464 and 4512~\AA\ in addition to the Mg \textsc{II} line.  The Ti abundance was allowed to vary, while the Fe abundance was fixed at the solar value. The Ti \textsc{II} lines near 4468~\AA\ were too strong in the model relative to the data, so we used only the line features between 4476 and 4512~\AA.  The best-fitting model had $V_{\mathrm{rot}}\sin i=100.9$ km s$^{-1}$, a Mg abundance of 5.2 times solar, and a Ti abundance of 5.6 times solar. As shown in Figure \ref{fig:mgb}, the fit is quite good.

	\subsubsection{Rotational Velocity from Broadening Kernel}\label{rotvelmodeling}
\begin{figure}[b]
	\epsscale{1.1}
	\plotone{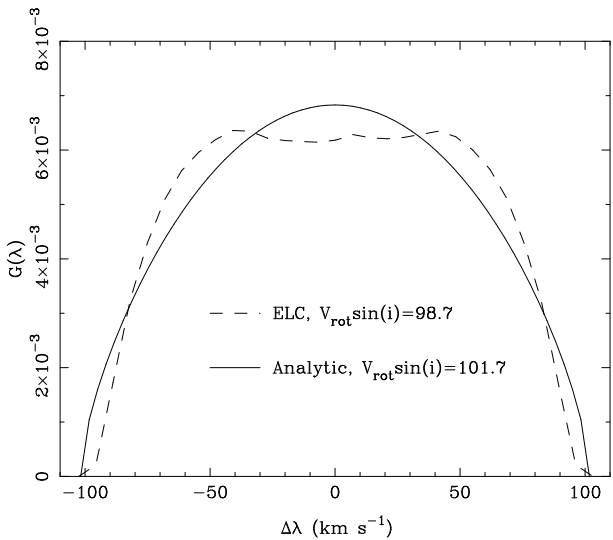}
	\caption{Analytic broadening kernel corresponding to a rotational velocity of 101.7 km s$^{-1}$ (solid line), which was measured from the Mg \textsc{II} line profile shown in Figure \protect\ref{fig:mgb}. The dashed line is the best fitting numerical kernel with a rotational velocity of 98.7 km s$^{-1}$, computed using the proper Roche geometry and at the phases of the observations.  \label{fig:plotkerns}}
\end{figure}

For a Roche lobe filling star, the broadening kernel can be distorted, particularly near phase 0.5, when the $L_{1}$ point comes in to full view (see Figure \ref{fig:plotkerns} for an example of this distortion). To analyze this distortion, we used ELC to compute numerical broadening kernels at the same phases as our observed spectra. Each broadening kernel was computed by averaging over the phases covered by the exposure time for each UVES spectrum. This gave us four numerical kernels matched to our observed spectra. These four kernels were combined to form an average numerical broadening kernel, in the same manner in which our four spectra were combined to give an average line profile. We generated an analytic broadening kernel using the rotational velocity measured from our spectra, 101.7 $\pm$ 0.8 km s$^{-1}$, then looked for the parameters for the numerical kernel which were the best fit to that analytic kernel. The best-fit rotational velocity from this procedure was $V_{\mathrm{rot}}\sin i=98.7\pm 0.8$ km s$^{-1}$, as shown in Figure \ref{fig:plotkerns}.

\subsection{Systematic Uncertainties}\label{sysuncert}

The ELC code produces final values for its fitted and calculated parameters with statistical uncertainties. These uncertainties are not the total uncertainties on the physical quantities which describe the system being investigated, they are simply a measure of how closely the model resembles the data. If the model resembles the data very closely, it is a good fit and the statistical uncertainties will be small. However, in some circumstances there may be a family of models, with comparable but not identical parameters, which give a similarly good fit to the data. In these cases, the uncertainties generated by ELC can be substantially less than the true uncertainties, and it is the range of parameter values in the model families which provides a reasonable measure of the underlying physical quantities and their uncertainties.

\citet{kreidberg_mass_2012} explore sources of systematic uncertainties in modeling X-ray binaries. They find that these systematics can be significant, particularly in systems with a faint companion star. If the brightness of the companion star is similar to or fainter than the brightness of the accretion disk, then changes in the disk emission can have a significant effect on the total emission from the system. A0620-00 is a good example of this, showing changes in both total brightness and the shape of the light curve which can be attributed to changes in the nonstellar light in the system \citep{Cantrell2010}. \citet{kreidberg_mass_2012} predict that these systematics should be small in a system such as V4641 Sgr, with its large bright companion star. However, since the basic parameter optimization using ELC only provides statistical uncertainties, we examined possible sources of systematic error in our modeling in order to provide more realistic uncertainties on the physical quantities of interest. 

In this section we describe in detail sources of systematic uncertainties and their effects on the final parameter determinations for V4641 Sgr. Those effects, as predicted by \citet{kreidberg_mass_2012}, are small here, but are included in the final calculations of uncertainties for the physical quantities of interest.

For the discussions which follow, we note that the output from ELC for a given parameter or physical quantity is actually a distribution of values. For all quantities discussed here, we calculated the mean and $\sigma$ (standard deviation) of the distribution of values, then we compared the mean to the value associated with the lowest $\chi^{2}$ value. In most cases these values were the same, and in the few cases where this was not true we used the parameter value from the model with the minimum $\chi^{2}$. Values shown in Table \ref{table:paramvariation} are this mean (or minimum $\chi^{2}$ value) and $\sigma$.

	\subsubsection{Uncertainties in Rotational Broadening}\label{sect:vrotsini}

In modeling V4641 Sgr, we had multiple reasonable values which could be used for the observed rotational velocity, as discussed in \S\ref{v4641inputs}. We therefore deemed it important to investigate the effects which different input velocities had on the output physical quantities. To do this we ran a series of models which only differed in their input for the observed $V_{\mathrm{rot}}\sin i$. Trial values used for $V_{\mathrm{rot}}\sin i$ and its uncertainty are shown in Table \ref{table:paramvariation}. We investigated six different values for $V_{\mathrm{rot}}\sin i$, taken from the literature, from our spectroscopy, and from some initial models we ran. 

As can be seen, varying the input rotational velocity had little effect on the final physical parameters derived from the model. 

While investigating this issue we also looked at changes in the outputs due to different levels of uncertainty in the rotational velocity. To do this we ran ELC using velocities of $95 \pm 1$ km s$^{-1}$ and of $124 \pm 1$ km s$^{-1}$. These two velocities were given in the literature with larger uncertainties, so reducing those to 1 km s$^{-1}$ each gave us a way to check this issue. We found that the models using $95 \pm 1$ km s$^{-1}$ gave results completely consistent with those in Table \ref{table:paramvariation}. We found that the models using $124 \pm 1$ km s$^{-1}$ produced somewhat different outputs: $Q = 1.93 \pm 0.12$, $M_{\mathrm{BH}} = 7.31 \pm 0.99$, and $M_{2} = 3.79 \pm 0.53$, showing that a higher precision can affect the modeling outcome if the velocity is significantly different from the velocity that is otherwise natural for observed data.

If a future spectroscopic study of this object produces a highly precise rotational velocity which is closer to 120 km s$^{-1}$ than to 100 km s$^{-1}$, then the physical quantities derived for this system will need to be revised. For now our best result is determined by an average of ELC's predictions of rotational velocities from all of these trials, $94.6 \pm 6.2$ km s$^{-1}$. This is consistent with the values determined from both high resolution spectroscopy and modeling.

	\subsubsection{Uncertainties in Light from Accretion Disk}\label{sect:diskfraction}

We believe V4641 Sgr has little to no light coming from the accretion disk in the passive state. However, light from the accretion disk can have a significant effect on the total emission from an X-ray binary, and 
data from our SMARTS spectra are consistent with a small fraction of the light, up to 15\%, coming from the disk. Therefore we wanted to  investigate the change in final physical quantities produced by modeling different values of light from the disk. We produced additional models with four trial values for the disk fraction present in the $V$ band, and with several levels of uncertainty for each of those disk fraction values. This allowed us to quantify the change in final physical parameters due to either failing to include light from the disk, or to over- or under-estimating the uncertainty in the disk fraction. Trial values tested for both the disk fraction and its uncertainty are listed in Table \ref{table:paramvariation}.

Our trials showed that changing the disk fraction, in a system which has a bright companion star and a faint accretion disk, does not appreciably affect the values for the calculated physical quantities. In fact in modeling V4641 Sgr, ELC consistently predicted no disk fraction for all trial input values.

However, although the effect of changing the disk fraction was very small for V4641 Sgr, it could be considerably enhanced in an XRB with a fainter companion star or a brighter disk. In these cases this source of systematic uncertainty would need to be treated carefully and thoroughly.

	\subsubsection{Uncertainties from Model Atmospheres}\label{sect:modelatmos}

In order to characterize the uncertainties associated with the assumed model atmospheres, we compared 
trials which used various analytic limb-darkening laws to trials we had already run using the detailed model atmospheres. We investigated three different limb-darkening laws: linear, log, and square-root. We ran trials in two different ways. For the first set we simply chose a limb-darkening law instead of the model atmospheres, but left all other inputs the same as previous trials which used model atmospheres.
For the second set, we compared model atmospheres and limb-darkening laws more directly, by using output from a modeling run as input to a second modeling run. That is, we took the output model curve from an ELC trial which used model atmospheres (as the curves shown in Figure \ref{fig:elcfits}), and gave that to ELC as the data to be fit (in effect using simulated data rather than the observed photometry). Thus we were able to test how closely a limb-darkening law could fit a model atmosphere.
Results from all trials are shown in Table \ref{table:paramvariation}. 

Using any analytic limb-darkening law to fit the observed data produced a somewhat lower inclination angle than using model atmospheres, leading to higher calculated masses for both the black hole and the companion star. The mass ratio and predicted output $V_{\mathrm{rot}}\sin i$ differ from those achieved using model atmospheres, some higher and some lower. Using a limb-darkening law to fit simulated data gave similar inclination angles but slightly different values for other parameters. 

We concluded that using these limb-darkening laws introduced a systematic uncertainty of approximately 10\% in the final calculated parameters. However, we believed this was an overestimation of the systematic errors associated with model atmosphere tables themselves. When calculating a model with analytic limb-darkening, ELC uses the same limb-darkening law for the entire star. This fails to account for the large range of temperatures found on the surface of the companion star due to gravity darkening, and also for strange effects introduced near the actual limb of the companion star due to the low gravity. Model atmosphere tables allow one to have location-specific limb-darkening information, accounting for drastic differences in temperature and gravity over the entire surface of the companion star. 

We determined that 5\% was a reasonable value for systematic uncertainties introduced by using model atmosphere tables, and we included this factor in our final calculations.

	\subsubsection{Uncertainties from Using Smaller Data Sets}\label{sect:datasubsets}

Data sets of significantly differing lengths, from hours to years, can be used to model binary systems.
Although a system can be modeled using as little as a single night of data, results from such studies can differ dramatically from the results achieved by using a much larger data sample \citep{kreidberg_mass_2012}. \citet{Cantrell2010}, for example, showed that using data sets of significantly different lengths in modeling A0620-00 can lead to changes of 20 degrees or more in the calculated inclination angle. This is an uncertainty which must be taken into account in these studies.

Our data set for V4641 Sgr is quite long, spanning approximately 10 years. However, in order to investigate the effect of different size data sets, we broke our data into subsets and ran ELC on each of those subsets individually. Each subset contained one passive period of data, generally covering several months. We required that each subset have an average of one data point every other night and that it be at least two months long, to assure that ELC would have enough data to constrain a model. Subsets used and outputs from the modeling are shown in Table \ref{table:paramvariation}. 

We found in this case that the final parameters for V4641 Sgr varied little when using smaller data sets. The subsets from late 2004, 2009, and 2010 produced slightly higher inclination angles, and therefore slightly lower component masses, but even with this all the results were consistent within the uncertainties.

We note that the possible effects from using fewer data can vary greatly from one XRB to another, but will certainly be larger in systems where the light curve changes shape or displays any other long-term trend in addition to the standard ellipsoidal variations.

\begin{deluxetable*}{p{2.8cm}ccccccc}
\tablewidth{0pt}
\tablecaption{Parameter variation trials \label{table:paramvariation}}
\tablehead{
\colhead{Trial} & 
\colhead{Inclination} & \colhead{Mass} & \colhead{} & \colhead{} & \colhead{Predicted} & \colhead{Predicted} & \colhead{Reduced} \\
\colhead{Parameter} & \colhead{Angle ($i$)} & \colhead{Ratio ($Q$)} & \colhead{$M_{\mathrm{BH}}$} & \colhead{$M_{\mathrm{2}}$} & \colhead{$V_{\mathrm{rot}}\sin i$} & \colhead{Disk Fraction} & \colhead{$\chi^{2}$} \\
\colhead{Values} & \colhead{(deg)} & \colhead{} & \colhead{(M$_{\sun}$)} & \colhead{(M$_{\sun}$)} & \colhead{(km s$^{-1}$)} & \colhead{($\times 10^{-8})$} & \colhead{}
}
\startdata
\sidehead{$V_{\mathrm{rot}}\sin i$}
124 $\pm$ 4\tablenotemark{a} & 72.52 $\pm$ 4.01 & 2.35 $\pm$ 0.29 & 6.37 $\pm$ 0.98 & 2.76 $\pm$ 0.56 & 91.57 $\pm$ 6.34 & $5.23 \pm 26.6$ & 0.99 \\
95 $\pm$ 10\tablenotemark{b} & 72.27 $\pm$ 3.93 & 2.30 $\pm$ 0.28 & 6.40 $\pm$ 0.97 & 2.84 $\pm$ 0.54 & 91.57 $\pm$ 6.18 & $5.16 \pm 27.9$ & 0.97 \\
101.7 $\pm$ 0.8\tablenotemark{c} & 72.50 $\pm$ 4.63 & 2.30 $\pm$ 0.18 & 6.48 $\pm$ 1.10 & 2.85 $\pm$ 0.54 & 93.07 $\pm$ 6.21 & $5.04 \pm 30.8$ & 1.00 \\
100.9 $\pm$ 0.8\tablenotemark{d} & 72.42 $\pm$ 4.63 & 2.32 $\pm$ 0.19 & 6.40 $\pm$ 1.07 & 2.81 $\pm$ 0.53 & 91.78 $\pm$ 6.08 & $5.40 \pm 28.0$ & 1.00 \\
98.7 $\pm$ 0.8\tablenotemark{e} & 72.31 $\pm$ 4.12 & 2.30 $\pm$ 0.27 & 6.39 $\pm$ 1.00 & 2.86 $\pm$ 0.54 & 91.58 $\pm$ 6.14 & $5.22 \pm 29.3$ & 0.99 \\
93.5 $\pm$ 1.0\tablenotemark{f} & 72.75 $\pm$ 3.65 & 2.35 $\pm$ 0.32 & 6.34 $\pm$ 0.91 & 2.73 $\pm$ 0.55 & 91.56 $\pm$ 6.24 & $5.06 \pm 24.3$ & 0.97 \\
\hline
\sidehead{Disk Fraction} 
0.05 $\pm$ 0.01 & 73.10 $\pm$ 4.42 & 2.11 $\pm$ 0.11 & 6.84 $\pm$ 0.97 & 3.25 $\pm$ 0.48 & 98.64 $\pm$ 5.67 & $ 4.61 \pm 25.4$ & 1.02 \\
0.10 $\pm$ 0.01 & 73.10 $\pm$ 4.39 & 2.10 $\pm$ 0.11 & 6.84 $\pm$ 0.95 & 3.25 $\pm$ 0.48 & 98.66 $\pm$ 5.62 & $ 4.62 \pm 25.0$ & 1.05 \\
0.15 $\pm$ 0.01 & 73.10 $\pm$ 4.48 & 2.10 $\pm$ 0.11 & 6.84 $\pm$ 0.97 & 3.25 $\pm$ 0.49 & 98.66 $\pm$ 5.74 & $ 4.66 \pm 25.3$ & 1.09 \\
0.20 $\pm$ 0.01 & 73.10 $\pm$ 4.42 & 2.11 $\pm$ 0.11 & 6.84 $\pm$ 1.00 & 3.25 $\pm$ 0.50 & 98.62 $\pm$ 5.79 & $ 4.60 \pm 26.6$ & 1.14 \\
0.05 $\pm$ 0.05 & 73.10 $\pm$ 4.33 & 2.10 $\pm$ 0.11 & 6.84 $\pm$ 0.98 & 3.25 $\pm$ 0.49 & 98.64 $\pm$ 5.69 & $ 4.61 \pm 26.5$ & 1.02 \\
0.10 $\pm$ 0.05 & 73.10 $\pm$ 4.33 & 2.10 $\pm$ 0.11 & 6.84 $\pm$ 0.98 & 3.25 $\pm$ 0.49 & 98.64 $\pm$ 5.69 & $ 4.61 \pm 26.5$ & 1.02 \\
0.15 $\pm$ 0.05 & 73.10 $\pm$ 4.33 & 2.10 $\pm$ 0.11 & 6.84 $\pm$ 0.98 & 3.25 $\pm$ 0.49 & 98.64 $\pm$ 5.69 & $ 4.61 \pm 26.5$ & 1.02 \\
0.20 $\pm$ 0.05 & 73.10 $\pm$ 4.33 & 2.10 $\pm$ 0.11 & 6.84 $\pm$ 0.98 & 3.25 $\pm$ 0.49 & 98.64 $\pm$ 5.69 & $ 4.61 \pm 26.5$ & 1.02 \\
0.05 $\pm$ 0.10 & 73.10 $\pm$ 4.40 & 2.10 $\pm$ 0.11 & 6.84 $\pm$ 0.95 & 3.25 $\pm$ 0.48 & 98.65 $\pm$ 5.58 & $ 4.61 \pm 24.9$ & 1.02 \\
0.10 $\pm$ 0.10 & 73.10 $\pm$ 4.33 & 2.10 $\pm$ 0.11 & 6.84 $\pm$ 0.98 & 3.25 $\pm$ 0.49 & 98.64 $\pm$ 5.69 & $ 4.61 \pm 26.5$ & 1.02 \\
0.15 $\pm$ 0.10 & 73.10 $\pm$ 4.33 & 2.10 $\pm$ 0.11 & 6.84 $\pm$ 0.98 & 3.25 $\pm$ 0.49 & 98.64 $\pm$ 5.69 & $ 4.61 \pm 26.5$ & 1.02 \\
0.20 $\pm$ 0.10 & 73.10 $\pm$ 4.33 & 2.10 $\pm$ 0.11 & 6.84 $\pm$ 0.98 & 3.25 $\pm$ 0.49 & 98.64 $\pm$ 5.69 & $ 4.61 \pm 26.5$ & 1.02 \\
\hline
\sidehead{Limb-Darkening Law}
Linear 

(passive photometry) & 68.90 $\pm$ 4.51 & 2.16 $\pm$ 0.14 & 7.29 $\pm$ 1.02 & 3.38 $\pm$ 0.50 & 97.55 $\pm$ 5.64 & $ 7.25 \pm 1.86$ & 0.98 \\
Log 

(passive photometry) & 68.16 $\pm$ 3.48 & 1.83 $\pm$ 0.09 & 8.18 $\pm$ 0.89 & 4.52 $\pm$ 0.51 & 106.52 $\pm$ 5.36 & $ 6.83 \pm 1.32$ & 0.97 \\
Square-root 

(passive photometry) & 68.33 $\pm$ 4.58 & 2.04 $\pm$ 0.11 & 7.64 $\pm$ 1.02 & 3.81 $\pm$ 0.47 & 100.66 $\pm$ 5.28 & $ 7.84 \pm 130$ & 0.99 \\
Linear 

(simulated data) & 72.83 $\pm$ 3.48 & 1.88 $\pm$ 0.11 & 7.40 $\pm$ 0.69 & 3.94 $\pm$ 0.40 & 105.31 $\pm$ 4.60 & $ 4.20 \pm 1.36$ & 0.48 \\
Log 

(simulated data) & 73.64 $\pm$ 2.56 & 2.12 $\pm$ 0.09 & 6.92 $\pm$ 0.59 & 3.17 $\pm$ 0.33 & 100.41 $\pm$ 3.64 & $ 3.89 \pm 0.97$ & 0.52 \\
Square-root 

(simulated data) & 72.16 $\pm$ 3.36 & 1.68 $\pm$ 0.10 & 7.86 $\pm$ 0.88 & 4.95 $\pm$ 0.62 & 109.87 $\pm$ 5.61 & $ 4.52 \pm 8.74$ & 0.55 \\
\hline
\sidehead{Data Subset}
2001 Mar 27 -- Oct 12 & 71.61 $\pm$ 3.62 & 2.11 $\pm$ 0.09 & 7.00 $\pm$ 0.81 & 3.32 $\pm$ 0.40 & 98.54 $\pm$ 4.69 & $5.34 \pm 29.4$ & 0.62 \\
2003 Apr 8 -- Jul 28 & 72.45 $\pm$ 3.51 & 2.11 $\pm$ 0.09 & 6.91 $\pm$ 0.78 & 3.25 $\pm$ 0.39 & 98.54 $\pm$ 4.52 & $4.43 \pm 20.3$ & 0.54 \\
2003 Aug 13 -- Nov 14 & 73.12 $\pm$ 3.44 & 2.11 $\pm$ 0.09 & 6.84 $\pm$ 0.77 & 3.25 $\pm$ 0.38 & 98.62 $\pm$ 4.44 & $3.54 \pm 27.7$ & 0.32 \\
2004 Mar 17 -- Jun 19 & 73.15 $\pm$ 3.44 & 2.11 $\pm$ 0.08 & 6.83 $\pm$ 0.77 & 3.25 $\pm$ 0.38 & 98.63 $\pm$ 4.47 & $4.30 \pm 29.4$ & 0.35 \\
2004 Jul 24 -- Oct 31 & 77.03 $\pm$ 3.94 & 2.11 $\pm$ 0.09 & 6.45 $\pm$ 0.81 & 3.30 $\pm$ 0.40 & 98.29 $\pm$ 4.57 & $4.48 \pm 19.6$ & 0.49 \\
2008 Feb 29 -- Sep 15 & 73.08 $\pm$ 3.44 & 2.11 $\pm$ 0.09 & 6.83 $\pm$ 0.76 & 3.24 $\pm$ 0.38 & 98.52 $\pm$ 4.46 & $4.70 \pm 22.7$ & 0.33 \\
2009 Mar 21 -- Sep 5 & 77.36 $\pm$ 3.70 & 2.10 $\pm$ 0.08 & 6.46 $\pm$ 0.78 & 3.10 $\pm$ 0.39 & 98.77 $\pm$ 4.41 & $3.30 \pm 16.6$ & 0.41 \\
2010 Feb 14 -- Jul 13 & 77.34 $\pm$ 3.74 & 2.11 $\pm$ 0.08 & 6.45 $\pm$ 0.79 & 3.08 $\pm$ 0.39 & 98.63 $\pm$ 4.46 & $3.29 \pm 13.4$ & 0.33 
\enddata
\tablecomments{Output from modeling is a distribution of values for each physical parameter; shown here are mean and 1-$\sigma$ for the distribution for that parameter in that model. In models where the mean value for a parameter was significantly different from that associated with the minimum $\chi^{2}$ value, we show here the value associated with the minimum $\chi^{2}$.
See \S\ref{elcdetails} for explanation of ``Predicted'' values.}
\tablenotetext{a}{Source: \citet{orosz_black_2001}}
\tablenotetext{b}{Source: \citet{sadakane}}
\tablenotetext{c}{Source: best-fitting model using Mg \textsc{II} alone [statistical uncertainty only]}
\tablenotetext{d}{Source: best-fitting model using Mg \textsc{II}, Ti \textsc{II}, and Fe \textsc{II} lines [statistical uncertainty only]}
\tablenotetext{e}{Source: ELC computation of broadening kernel [statistical uncertainty only]}
\tablenotetext{f}{Source: prediction of one ELC model (at low end of range of velocities) [statistical uncertainty only]}
\end{deluxetable*}

\subsection{Our Final Parameters \& Uncertainties}\label{finalparamvalues}

Taking into account all of the factors discussed in sections \ref{v4641inputs} and \ref{sysuncert}, we arrived at the final input observed values of the rotational velocity and the disk fraction for modeling V4641 Sgr.

Varying the input $V_{\mathrm{rot}}\sin i$ gave an average predicted $V_{\mathrm{rot}}\sin i$ of $93.3 \pm 6.2$ km s$^{-1}$. Varying the type of atmosphere calculation gave an average predicted $V_{\mathrm{rot}}\sin i$ of $103.4 \pm 5.0$ km s$^{-1}$. Varying the input disk fraction gave an average predicted $V_{\mathrm{rot}}\sin i$ of $98.6 \pm 5.7$ km s$^{-1}$, while modeling only smaller sets of data gave $98.6 \pm 4.5$ km s$^{-1}$. To account for all of these results we took the weighted average of these, $V_{\mathrm{rot}}\sin i = 98.5 \pm 5.3$ km s$^{-1}$, as the appropriate input value for modeling.

All of our parameter variation consistently gave a result of no predicted disk fraction, therefore we included no light from the accretion disk in our final modeling.

Using these values, and those given earlier in Table \ref{table:inputelcparams}, we computed a final group of ELC models to obtain physical quantities which reasonably described this system. These all had the same six free parameters, listed in a different order in the input files for each modeling run.

Our final best-fit parameters (also shown in Table \ref{table:finalparams}) are: 
$i = 72.3 \pm 4.1\degr$, 
$Q = 2.2 \pm 0.2$, 
$a = 17.5 \pm 1.0$ R$_{\sun}$, 
$r_{\mathrm{out}} = 0.7 \pm 0.1$, 
$\beta_{\mathrm{rim}} = 0.1 \pm 0.2\degr$, 
$T_{2} = 10250 \pm 300$ K, 
$R_{2} = 5.3 \pm 0.3$ R$_{\sun}$, 
$M_{\mathrm{BH}} = 6.4 \pm 0.6$ M$_{\sun}$, 
and 
$M_{2} = 2.9 \pm 0.4$ M$_{\sun}$. 
The uncertainties quoted include both the statistical uncertainties from ELC itself and the systematic uncertainties we calculated in the previous subsections.

\begin{deluxetable}{cc}
\tablewidth{0pt}
\tablecaption{Best-fit model parameters for V4641 Sgr, using passive state data\label{table:finalparams}}
\tablehead{\colhead{Parameter} & \colhead{Best-fit Value} }
\startdata
$i$ & $72.3 \pm 4.1\degr$ \\
$Q$ & $2.2 \pm 0.2$ \\
$a$ & $17.5 \pm 1.0$ R$_{\sun}$ \\
$r_{\mathrm{out}}$ & $0.7 \pm 0.1$ \\
$\beta_{\mathrm{rim}}$ & $0.1 \pm 0.2\degr$ \\
$T_{2}$ & $10250 \pm 300$ K \\
$R_{2}$ & $5.3 \pm 0.3$ R$_{\sun}$ \\
$M_{\mathrm{BH}}$ & $6.4 \pm 0.6$ M$_{\sun}$ \\
$M_{2}$ & $2.9 \pm 0.4$ M$_{\sun}$ \\
distance & $6.2 \pm 0.7$ kpc \\
\hline \\[-6pt]
Red. $\chi^{2}$ & 0.97
\enddata
\end{deluxetable}

With these new parameters we were able to calculate an updated distance to V4641 Sgr. We used the cross-section of the companion star at secondary minimum (when it most resembles a spherical star) to calculate an absolute $V$ magnitude, and the reddening and apparent $V$ magnitude to convert that to a distance estimate. We calculated the distance to be $6.2 \pm 0.7$ kpc.

\begin{figure*}[tb]
	\epsscale{1.0}
	\plotone{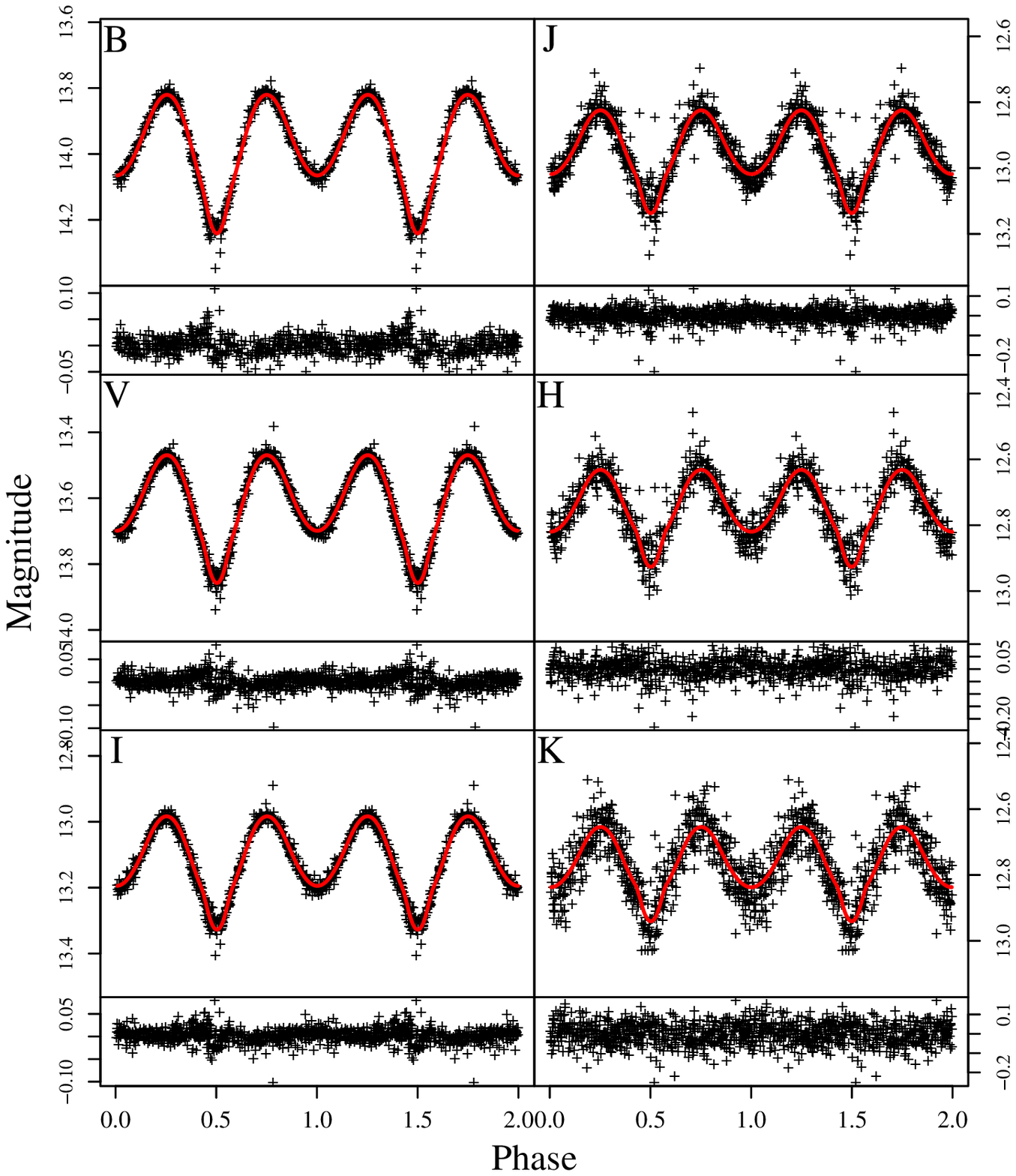}
	\caption{Ellipsoidal Light Curve (ELC) model fit to the passive data for V4641 Sgr. Best-fit model parameters are given in Table \ref{table:finalparams}. Passive data are shown by plus symbols, model fits are thick curves (red in the online version). Beneath each panel is a plot of the residuals (data minus model). We emphasize that the curves in these plots are models of a physical system, generated to fit our data, in contrast to the passive envelope curve drawn in Figure \ref{fig:v4641-vpassiveenv}, which was empirical and drawn directly from the data. 
	Notice that even with these detailed models, we still do not completely match the data at phase 0.5. Again, we wonder if the extreme faintness of these points could be caused by an infrequent slight eclipse of the companion star by the colder accretion disk.
	\label{fig:elcfits}}
\end{figure*}

The model light curves obtained with these best-fit parameters are shown in Figure \ref{fig:elcfits}, overplotted on the passive-state data. Residuals from subtracting the model from the data are also shown for each band. We emphasize that the curves in these plots are models of a physical system, which were generated to fit our data, in contrast to the passive envelope drawn in Figure \ref{fig:v4641-vpassiveenv}, which was empirical and drawn directly from the data. 

\citet{orosz_black_2001} modeled V4641 Sgr using the same methodology and arrived at component masses and a distance which were significantly larger than those presented here: $8.73 \le M_{\mathrm{BH}} \le 11.70$ M$_{\sun}$, $5.49 \le M_{2} \le 8.14$ M$_{\sun}$, $7.40 \le d \le 12.31$ kpc. This can be explained by the particular parameters which were different from that study to this work. \citet{orosz_black_2001} used a rotational velocity of $124 \pm 4$ km s$^{-1}$, a mass ratio $Q$ of 1.5, and a range for the inclination angle of $60\degr - 70.7\degr$. A lower rotational velocity leads to a larger $Q$, and a larger $Q$ leads to a smaller final mass. Similarly, a larger inclination angle leads to a smaller final mass. Since the mass actually depends on the square of $Q$ and the $\sin^3(i)$, even small changes in $Q$ or $i$ lead to large changes in the final mass. Therefore our masses are smaller than those in the previous work. To calculate the distance to V4641 Sgr, \citet{orosz_black_2001} used $E(\bv) = 0.32 \pm 0.1$ and a companion star radius of $7.5 \pm 0.5$ R$_{\sun}$. Our reddening value was higher, and our companion star size was smaller, both of which effects led to a smaller calculated distance.

\section{Discussion \& Conclusions}\label{discussconcludesect}

We have examined $\sim$10 years of photometric data taken during X-ray quiescence in order to 
investigate variation in optical and infrared emission and to calculate more accurate masses for both the companion star and the black hole. We found that this source does have two separate states of behavior during quiescence: passive and active. The passive state is dominated by ellipsoidal variations, and is remarkably stable in the overall shape and variability characteristics of the light curve. The active state is both brighter and more variable than the passive state. These optical/infrared states last for weeks or months --- long compared to the orbital period of the system. In our data, V4641 Sgr spends approximately 85\% of X-ray quiescence in the passive state and approximately 15\% in the active state.

Analysis of the passive colors of V4641 Sgr shows that they are consistent with the emission from a solitary reddened B9III star with a reddening value of $E(\bv) = 0.37 \pm 0.19$. Spectroscopy indicates little to no emission from the accretion disk during the passive state. This contrasts with A0620-00, and other systems, where there was significant disk light contributing to the total emission during all states \citep{Cantrell2010, filippenko_mass_1995}.

The active state was studied by subtracting the passive envelope from the data in each band and examining the residuals, or ``excess light''. This active state excess light was found to change over the course of the orbital period, appearing brighter at phase 0.5 (when the accretion disk is between the observer and the companion star). We also found that the excess light during active states is redder than the companion star, in contrast to A0620-00. 

We used our long baseline of passive-state data to model the ellipsoidal variations in V4641 Sgr in order to determine the binary parameters of the system. We characterized major sources of systematic uncertainties in this type of modeling, including input observable quantities such as rotational velocity, the possibility of a fraction of light coming from the accretion disk, the choice of model atmospheres over analytic limb-darkening laws, and the use of smaller subsets of data. We incorporated all of these factors in the calculations of our final best-fit parameters.

We found that V4641 Sgr has an inclination angle of $i = 72.3\degr$, a mass ratio of $Q = 2.2$, component masses for the system of $M_{BH} =  6.4$ M$_{\sun}$ and $M_{2} = 2.9$ M$_{\sun}$, and is located at a distance of 6.2 kpc. 
Our best-fitting model does not predict an X-ray eclipse. X-ray observations are sparse near the phase when an eclipse would occur, however they do allow us to place an upper limit on eclipse duration of 8.3$\degr$ in phase, or $\sim$1.6 hours. 

Differing states of behavior in the optical and infrared may be common among quiescent black hole X-ray binaries. Understanding these states is key to gaining more accurate and more precise measurements for the binary parameters, and especially the component masses, in all of these systems. The active state may also be key in determining the nature of the quiescent accretion flow and in testing the existence of the event horizon \citep{narayan_advection_2008}. We look forward to future studies of other XRBs during X-ray quiescence to continue to expand our understanding of this state and its accretion flows.

\acknowledgements
We thank the anonymous referee for many useful comments which improved the quality of this paper.
RKDM and CDB received support from NSF grant AST-0707627.
This paper has made use of up-to-date SMARTS optical/near-infrared light curves that are available at www.astro.yale.edu/smarts/xrb/home.php. The Yale SMARTS XRB team is supported by NSF grants 0407063 and 070707 to Charles Bailyn.
This publication makes use of data products from the Two Micron All Sky Survey, which is a joint project of the University of Massachusetts and the Infrared Processing and Analysis Center/California Institute of Technology, funded by the National Aeronautics and Space Administration and the National Science Foundation.
This research has made use of the USNOFS Image and Catalogue Archive operated by the United States Naval Observatory, Flagstaff Station (http://www.nofs.navy.mil/data/fchpix/).
This research has made use of data obtained through the High Energy Astrophysics Science Archive Research Center Online Service, provided by the NASA/Goddard Space Flight Center.

{\it Facilities:} \facility{CTIO:SMARTS 1.3m}, \facility{CTIO:SMARTS 1.5m}, \facility{VLT:Kueyen (UVES)}, \facility{RXTE}

\bibliographystyle{apj}

\end{document}